\documentclass[article,reprint,superscriptaddress,longbibliography]{revtex4-1}
\usepackage{graphicx}
\usepackage{dcolumn}
\usepackage{bm}
\usepackage{hyperref} 
\usepackage{breakurl}
\usepackage{multirow}
\usepackage{enumitem}
\usepackage{color}
\usepackage[pagewise]{lineno}
\setcitestyle{super} 

\begin{document}

\title{Interfering Josephson diode effect in Ta$_{2}$Pd$_{3}$Te$_{5}$ asymmetric edge interferometer}

\author{Yupeng Li}
      \thanks{Equal contributions}
      \affiliation{Beijing National Laboratory for Condensed Matter Physics, Institute of Physics, Chinese Academy of Sciences, Beijing 100190, China}

\author{Dayu Yan}
      \thanks{Equal contributions}
      \affiliation{Beijing National Laboratory for Condensed Matter Physics, Institute of Physics, Chinese Academy of Sciences, Beijing 100190, China}

\author{Yu Hong}
      \affiliation{Beijing National Laboratory for Condensed Matter Physics, Institute of Physics, Chinese Academy of Sciences, Beijing 100190, China}
      \affiliation{School of Physical Sciences, University of Chinese Academy of Sciences, Beijing 100049, China}

\author{Haohao Sheng}
      \affiliation{Beijing National Laboratory for Condensed Matter Physics, Institute of Physics, Chinese Academy of Sciences, Beijing 100190, China}
      \affiliation{School of Physical Sciences, University of Chinese Academy of Sciences, Beijing 100049, China}

\author{Anqi Wang}
      \affiliation{Beijing National Laboratory for Condensed Matter Physics, Institute of Physics, Chinese Academy of Sciences, Beijing 100190, China}
      \affiliation{School of Physical Sciences, University of Chinese Academy of Sciences, Beijing 100049, China}

\author{Ziwei Dou}
      \affiliation{Beijing National Laboratory for Condensed Matter Physics, Institute of Physics, Chinese Academy of Sciences, Beijing 100190, China}

\author{Xingchen Guo}
      \affiliation{Beijing National Laboratory for Condensed Matter Physics, Institute of Physics, Chinese Academy of Sciences, Beijing 100190, China}
      \affiliation{School of Physical Sciences, University of Chinese Academy of Sciences, Beijing 100049, China}

\author{Xiaofan Shi}
      \affiliation{Beijing National Laboratory for Condensed Matter Physics, Institute of Physics, Chinese Academy of Sciences, Beijing 100190, China}
      \affiliation{School of Physical Sciences, University of Chinese Academy of Sciences, Beijing 100049, China}

\author{Zikang Su}
      \affiliation{Beijing National Laboratory for Condensed Matter Physics, Institute of Physics, Chinese Academy of Sciences, Beijing 100190, China}
      \affiliation{School of Physical Sciences, University of Chinese Academy of Sciences, Beijing 100049, China}

\author{Zhaozheng Lyu}
      \affiliation{Beijing National Laboratory for Condensed Matter Physics, Institute of Physics, Chinese Academy of Sciences, Beijing 100190, China}

\author{Tian Qian}
      \affiliation{Beijing National Laboratory for Condensed Matter Physics, Institute of Physics, Chinese Academy of Sciences, Beijing 100190, China}
      \affiliation{Songshan Lake Materials Laboratory, Dongguan 523808, China}

\author{Guangtong Liu}
      \affiliation{Beijing National Laboratory for Condensed Matter Physics, Institute of Physics, Chinese Academy of Sciences, Beijing 100190, China}
      \affiliation{Songshan Lake Materials Laboratory, Dongguan 523808, China}

\author{Fanming Qu}
      \affiliation{Beijing National Laboratory for Condensed Matter Physics, Institute of Physics, Chinese Academy of Sciences, Beijing 100190, China}
      \affiliation{School of Physical Sciences, University of Chinese Academy of Sciences, Beijing 100049, China}
      \affiliation{Songshan Lake Materials Laboratory, Dongguan 523808, China}

\author{Kun Jiang}
      \affiliation{Beijing National Laboratory for Condensed Matter Physics, Institute of Physics, Chinese Academy of Sciences, Beijing 100190, China}

\author{Zhijun Wang}
      \affiliation{Beijing National Laboratory for Condensed Matter Physics, Institute of Physics, Chinese Academy of Sciences, Beijing 100190, China}
      \affiliation{School of Physical Sciences, University of Chinese Academy of Sciences, Beijing 100049, China}

\author{Youguo Shi}
      \email{ygshi@iphy.ac.cn}
      \affiliation{Beijing National Laboratory for Condensed Matter Physics, Institute of Physics, Chinese Academy of Sciences, Beijing 100190, China}
      \affiliation{Songshan Lake Materials Laboratory, Dongguan 523808, China}

\author{Zhu-An Xu}
      \affiliation{School of Physics, Zhejiang University, Hangzhou, China}
      \affiliation{State Key Laboratory of Silicon and Advanced Semiconductor Materials, Zhejiang University, Hangzhou, China}
      \affiliation{Hefei National Laboratory, Hefei 230088, China}

\author{Jiangping Hu}
      \affiliation{Beijing National Laboratory for Condensed Matter Physics, Institute of Physics, Chinese Academy of Sciences, Beijing 100190, China}
      \affiliation{Kavli Institute of Theoretical Sciences, University of Chinese Academy of Sciences, Beijing 100190, China}

\author{Li Lu}
      \email{lilu@iphy.ac.cn}
      \affiliation{Beijing National Laboratory for Condensed Matter Physics, Institute of Physics, Chinese Academy of Sciences, Beijing 100190, China}
      \affiliation{School of Physical Sciences, University of Chinese Academy of Sciences, Beijing 100049, China}
      \affiliation{Songshan Lake Materials Laboratory, Dongguan 523808, China}

\author{Jie Shen}
      \email{shenjie@iphy.ac.cn}
      \affiliation{Beijing National Laboratory for Condensed Matter Physics, Institute of Physics, Chinese Academy of Sciences, Beijing 100190, China}
      \affiliation{Songshan Lake Materials Laboratory, Dongguan 523808, China}
      \affiliation{Beijing Academy of Quantum Information Sciences, Beijing 100193, China}

\begin{abstract}
\textbf{
Edge states in topological systems have attracted great interest due to their robustness and linear dispersions.
Here a superconducting-proximitized edge interferometer is engineered on
a topological insulator Ta$_{2}$Pd$_{3}$Te$_{5}$ with asymmetric edges to
realize the interfering Josephson diode effect (JDE), which hosts many advantages,
such as the high efficiency as much as 73\% at tiny applied magnetic fields with
an ultra-low switching power around picowatt.
As an important element to induce such JDE, the second-order harmonic
in the current-phase relation is also experimentally confirmed by half-integer Shapiro steps.
The interfering JDE is also accompanied by the antisymmetric second harmonic transport,
which indicates the corresponding asymmetry in the interferometer, as well as the polarity of JDE.
This edge interferometer offers an effective method to enhance the
performance of JDE, and boosts great potential applications for future superconducting quantum devices.
}
\end{abstract}

\maketitle
\noindent\textbf{Introduction}

The semiconductor diode is a fundamental component in modern electronics due to the non-reciprocal responses \cite{PNjunction_shockley_BSTJ1949}.
Analogous non-reciprocal charge transport in superconductors -
namely superconducting diode effect (SDE) - has great potential for
superconducting quantum electronics, since Josephson junctions (JJs)
and superconducting quantum interference devices
(SQUIDs) have been key components of superconducting quantum devices \cite{SQC2_YuY_science2002,SQC3_IoffeLB_Nature1999}.
SDE in the JJs - namely Josephson diode effect (JDE) - and in intrinsic superconductors has been theoretically proposed
in various systems with broken time-reversal and inversion symmetries
\cite{JDE_HJP_PRL2007,AsymmetricJDE_ChenCZ_PRB2018,JDE_MisakiKou_PRB2021,ISDE_Daido_PRL2022,SDEFMS_YuanNFQ_PNAS2022,
FDEinSI_Constanti_PRL2022,JDE_FuLiang_SA2022,GeneralJDE_PRX2022,SDE_HJJ_NJP2022}.
Field-induced and field-free SDE/JDE has been experimentally observed in various superconductors
\cite{NbTaVSDE_AndoF_nature2020,NiTe2JDE_PalB_NP2022,ThreeGrapheneFFSDE_LinJXZ_NP2022,Y3Fe5O12FFJDE_JeonKR_NM2022,
NbSe2SCE_BauriedlL_NC2022,JDEintwistBilayerGra_NC2023,InSbFlakeJDE_NanoL2022,NbVCoVTaFFSDE_NN2022},
supercurrent interferometers \cite{InAsSQUID_arXiv2023,TiSQUIDSDE_PaolucciF_APL2023,JDEInGaAs3terminal_NC2023,2DgasSQUIDdiode_arXiv2023},
and other systems \cite{Nb3Cl8FFSDE_WH_Nature2022,FieldfreeSDE_NC2022,VFilm_HouYS_arXiv2022,JDEmageticAtton_TrahmsM_nature2023,
DiamagneticSDE_SundareshA_NC2023}.
Interestingly, asymmetric supercurrent interferometers or SQUIDs with a non-sinusoidal current-phase relation (CPR)
provide a good platform to realize JDE (Fig. 1a) with ultra-small magnetic fields and ultra-low power consumption \cite{TiSQUIDSDE_PaolucciF_APL2023,InAsSQUID_arXiv2023},
both of which are crucial elements for applications at ultra-low temperatures.
Moreover, it is quite feasible to promote their efficiency by varying the generic configuration \cite{FDEinSI_Constanti_PRL2022},
which is compatible with state-of-the-art lithography technology,
and integrate them into large-scale superconducting quantum circuits.

\begin{figure*}[!thb]
\begin{center}
\includegraphics[width=7in]{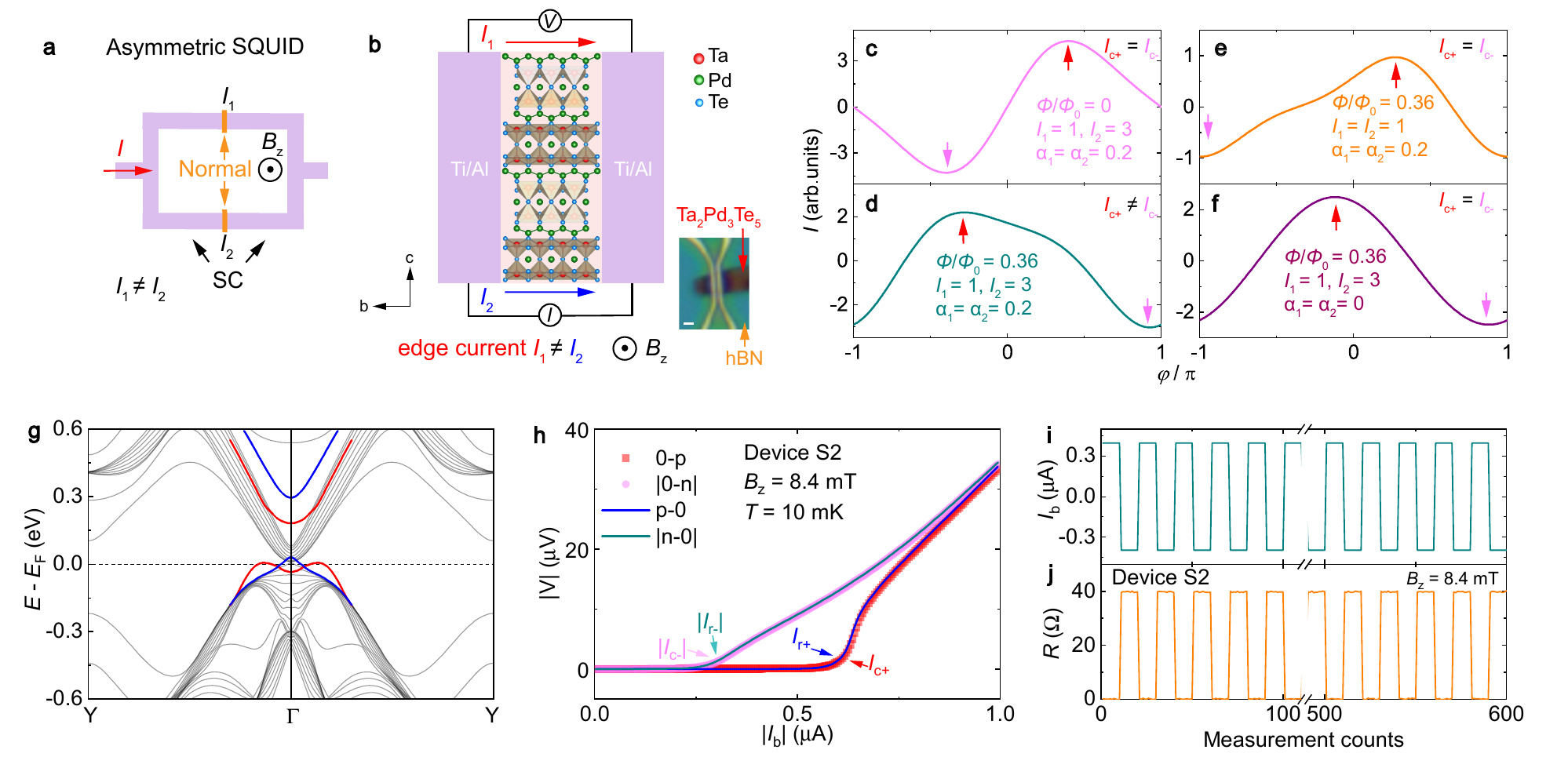}
\end{center}
\caption{\label{Fig1} \textbf{Mechanisms for SDE/JDE.}
$\mathbf{a}$, JDE in the asymmetric SQUID formed by two asymmetric JJs.
$\mathbf{b}$, An asymmetric SQUID formed by asymmetric edge states in Ta$_{2}$Pd$_{3}$Te$_{5}$ JJ
and its device configuration. The upper and lower edges are the palladium and tantalum-tellurium atomic chain, respectively.
The inset shows the photomicrograph of device S2, and the white scale bar corresponds to 1 $\mu$m.
$\mathbf{c-f}$, Simulated CPR based on the minimal model.
JDE only exists in $\mathbf{d}$ with different supercurrents, a non-zero
magnetic field ($\phi/\phi_{0} \neq 0$) and higher harmonics in the CPR ($\alpha_{n} \neq 0$).
The red and pink arrows represent the maximum($I_{c+}$) and minimum($I_{c-}$) current in the CPR, respectively.
$\mathbf{g}$, Band structure of monolayer Ta$_{2}$Pd$_{3}$Te$_{5}$.
Edge states (blue and red lines) exist near the Fermi level
and are marked in the corresponding edges of $\mathbf{b}$.
$\mathbf{h}$, $\left|I_{b}\right|-\left|V\right|$ curves for positive
and negative current sweep at $B_{z}$ = 8.4 mT.
$\mathbf{i,j}$, Alternating switching between the superconducting and normal states at $B_{z}$ = 8.4 mT and 10 mK in device S2.
   }
\end{figure*}

Essentially, the appearance of JDE is related to the difference between
maximum($I_{c+}$) and minimum($I_{c-}$) current in the CPR.
The conventional CPR is an odd function of superconducting phase difference $\varphi$ between superconducting leads,
entailing a zero supercurrent at $\varphi$ = 0,
which shifts [meaning $I(\varphi=0)$ $\neq$ 0] when both time-reversal and chiral symmetries are broken \cite{Phi0sysmmetry_np2016,InAsSQUID_arXiv2023,AJEinNW_PRB2014}.
To achieve JDE, other mechanisms to cause the CPR deviating from standard sinusoidal
$I(\varphi)$ = $I_{c}$ $sin(\varphi)$ will also be introduced.
The characteristics of CPR, which are usually inferred by the interference pattern of SQUIDs and
the Shapiro steps due to ac Josephson effect, as well as the newly-discovered JDE with non-reciprocal critical current,
have been widely harnessed to detect the anomalous superconductivity \cite{4piPeriodJJ_NC2016,EdgeSQUID_np2014},
such as proximitized helical/chiral topological edge states.
In turn, in a SQUID constructed by two asymmetric edge states of the topological insulator,
JDE with an unconventional CPR could be easily realized with highly-tunable efficiency \cite{AsymmetricJDE_ChenCZ_PRB2018} and relatively easy fabrication (Fig. 1b).
The relatively small number of edge supercurrent channels, in comparison to the case of bulk states transport,
may contribute to the small critical current and power consumption \cite{EdgeSQUID_np2014,EdgeSC_NN2015}.

More precisely, the interferometer formed by two asymmetric edge supercurrents (Fig. 1b)
can be viewed as a SQUID with two different JJs,
and induces JDE with requirements similar to the asymmetric SQUID \cite{FDEinSI_Constanti_PRL2022} in Fig. 1a.
According to the Souto $et$ $al$. \cite{FDEinSI_Constanti_PRL2022},
a minimal model with $n$ JJs concatenated in an interferometer array is applied to account for the origin of JDE,
and the CPR only takes the first and second harmonic contribution into consideration.
The current is written as $I (\varphi,\phi/\phi_{0})  = \sum_{1}^{n=2} I_{n}sin(\varphi + 2\pi(n-1)\phi/\phi_{0}) + \alpha_{n}I_{n}sin(2\varphi + 4\pi(n-1)\phi/\phi_{0})$ for two JJs in parallel,
where $I_{n}$ ($\alpha_{n}I_{n}$) is the amplitude of the first (second) harmonic content for the $n$th JJ,
$\alpha_{n}$ is current coefficient of the higher harmonic, $\phi$ is the magnetic flux
and $\phi_{0} = h/2e$ is the magnetic flux quantum.
In Fig. 1c-f, the simulations of the CPR using the aforementioned formula serves the purpose of illustrating three key requirements that contribute to the difference between $I_{c+}$ (red arrow) and $I_{c-}$ (pink arrow).
First, an external magnetic field ($\phi/\phi_{0} \neq 0$) is necessary to break time-reversal symmetry
and induce nonzero current at $\varphi$ = 0 in the CPR (Fig. 1d-f).
Second, the supercurrent of each edge, including all the harmonic contributions, should be different
(see Fig. 1d,e and detailed simulations in Supplementary Fig. 2).
This condition can be realized by two interfering edge states with different dispersion \cite{AsymmetricJDE_ChenCZ_PRB2018}.
It can also be met, for example, by two superconducting quantum point contacts
with different transmission coefficients \cite{FDEinSI_Constanti_PRL2022},
which may be linked to various factors such as different edge states, disorder situations, among others.
All of these can cause the different supercurrent of edges in our devices,
which is actually quite unique in topological materials.
Finally, at least one edge channel should be transmissive so that the CPR acquires
a higher harmonic ($\alpha_{n} \neq 0$) \cite{FDEinSI_Constanti_PRL2022,JDESQUID_Fominov_PRB2022}
(see Fig. 1d, f).

\vspace{3ex}
\noindent\textbf{Results}

\noindent\textbf{Interfering JDE}

In this work, highly efficient JDE, residing in the tilting supercurrent pattern,
is reported in a Ta$_{2}$Pd$_{3}$Te$_{5}$ edge interferometer at very small
$B_{z}$ and is driven by low power consumption to achieve a highly stable rectification effect.
It is also accompanied by fractional Shapiro steps, revealing the higher harmonic contributions of
the non-negligible transmission channels.
The choice of Ta$_{2}$Pd$_{3}$Te$_{5}$ is based on the following reasons:
(1) Ta$_{2}$Pd$_{3}$Te$_{5}$ is a van der Waals material with quasi-one-dimensional (1D) chains
\cite{Ta2Pd3Te5QSH_GZP_PRB21} and can be easily mechanically exfoliated.
(2) As a 2D topological insulator,
where multiple layers could be viewed as simple stacking of monolayers due to very weak inter-layer coupling
\cite{Ta2Pd3Te5TI_GuoZP_npjqm2022,Ta2Pd3Te5LL_arXiv2022}, it hosts excitonic insulator states \cite{Ta2Pd3Te5ExcitionIn_PRX2024,Ta2Pd3Te5ExcitionEdge_arXiv2023,Ta2Pd3Te5ExcitionIn2_PRX2024} and edge states
observed by scanning tunnelling microscopy \cite{Ta2Pd3Te5STM_WangXG_PRB2021,Ta2Pd3Te5ExcitionEdge_arXiv2023}
and transport measurements \cite{Ta2Pd3Te5LL_arXiv2022,Ta2Pd3Te5thermometer_Arxiv24}.
Notably, owing to the anisotropic bonding energy of 1D chains,
the edges possess different atomic chains (Pd and Ta-Te atomic chains), resulting in asymmetric dispersion,
as shown by the red and blue lines/arrows in Fig. 1b, g.
(3) Superconductivity can be easily achieved in this system through doping or high-pressure techniques
\cite{Ta2Pd3Te5dopingSC_JPSJ2021,Ta2Pd3Te5presureSC_arXiv2023,Ta2Ni3Te5pressureSC_YangHY_PRB2023}.
Therefore, by proximity effect, the induced supercurrent (Supplementary Fig. 1)
carried by edge states of Ta$_{2}$Pd$_{3}$Te$_{5}$ can be used to realize an asymmetric SQUID with the interfering JDE.

\begin{figure*}[!thb]
\begin{center}
\includegraphics[width=7in]{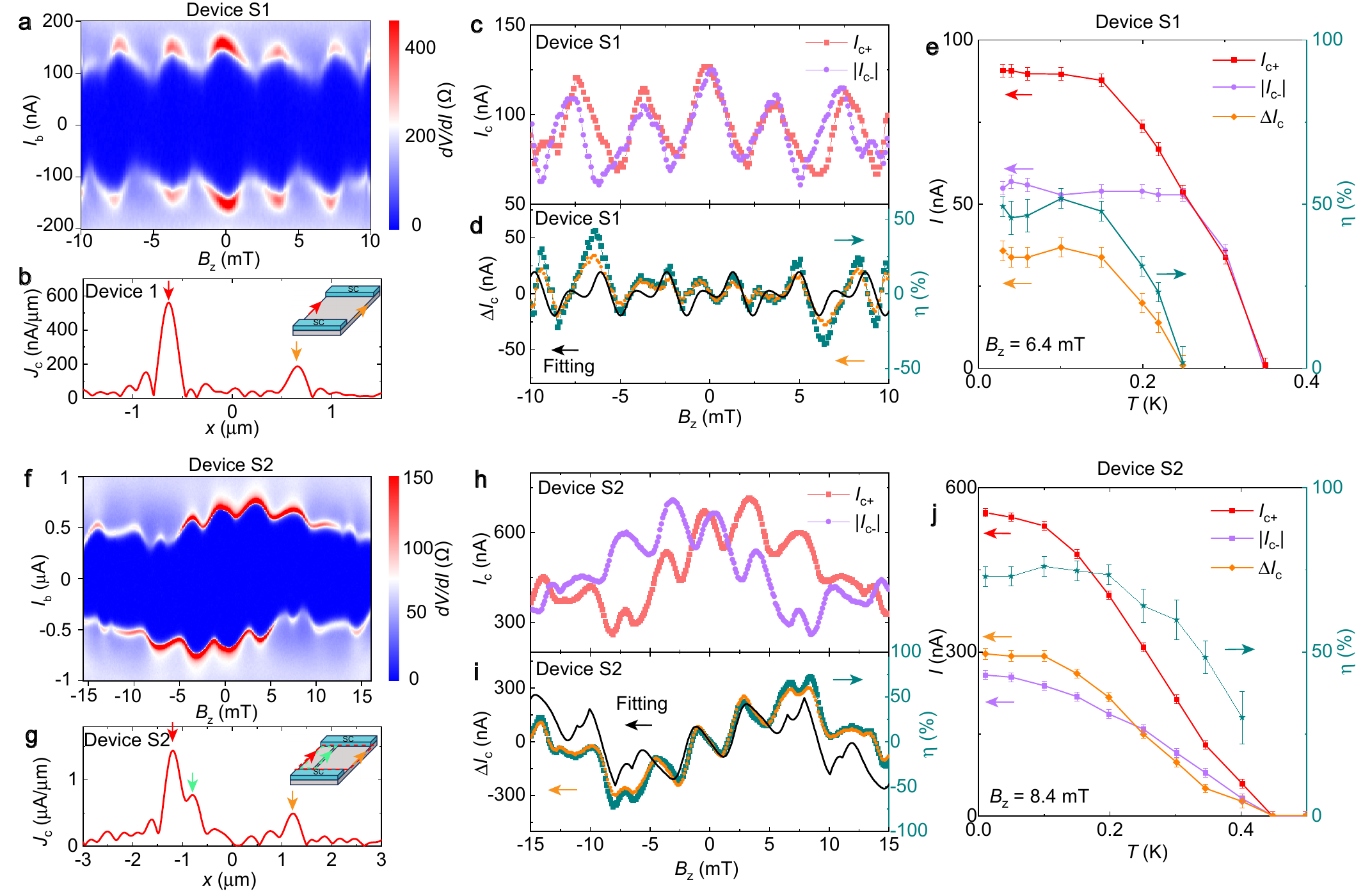}
\end{center}
\caption{\label{Fig2} \textbf{Asymmetric Josephson effect and JDE in Ta$_{2}$Pd$_{3}$Te$_{5}$ JJs.}
$\mathbf{a}$, SQUID pattern of device S1 at 15 mK.
$\mathbf{b}$, Position-dependent supercurrent density distribution.
Inset: A sketch of edge-supercurrents-formed JJ.
$\mathbf{c}$, $B_{z}$-dependent $I_{c+}$ and $\left|I_{c-}\right|$.
The switching current $I_{c+}$ and $I_{c-}$ are extracted by $\sim$ 15\% of normal resistance.
$\mathbf{d}$, Non-reciprocal critical current $\Delta I_{c} = I_{c+} - \left|I_{c-}\right|$,
Josephson diode efficiency $\eta$ and fitted $\Delta I_{c}$ (black line) as a function of $B_{z}$.
$\mathbf{e}$, Temperature-dependent $I_{c+}$, $I_{c-}$, $\Delta I_{c}$ and $\eta$ for device S1 at $B_{z}$ = 6.4 mT.
The error bars primarily stem from the definition of the switching current,
which is determined by 15\% $\pm$ 5\% drop of the normal resistance.
$\mathbf{f}$, SQUID pattern of device S2 at 10 mK.
$\mathbf{g}$, Supercurrent density distribution.
The inset is a draft of the JJ with three edge supercurrents.
$\mathbf{h}$, $B_{z}$ dependence of $I_{c+}$ and $\left|I_{c-}\right|$.
$\mathbf{i}$, Oscillating $\Delta I_{c}$, $\eta$ and fitted $\Delta I_{c}$ (black line).
$\mathbf{j}$, Temperature-dependent $I_{c+}$, $I_{c-}$, $\Delta I_{c}$ and $\eta$ for device S2 at $B_{z}$ = 8.4 mT.
The error bars are obtained by the same method as in $\mathbf{e}$.
 }
\end{figure*}

In Fig. 1h, JDE reaches up to 73\% efficiency
[$\eta$ = 2($I_{c+} - \left|I_{c-}\right|)/(I_{c+} + \left|I_{c-}\right|)$]
at an out-of-plane magnetic field $B_{z}$ = 8.4 mT in $\left|I_{b}\right|-\left|V\right|$ traces.
The measurement is performed in the following sequence:
(1) Zero-to-positive current sweep (0-p, red scatters);
(2) Positive-to-zero current sweep (p-0, blue line);
(3) Zero-to-negative current sweep (0-n, magenta scatters);
(4) Negative-to-zero current sweep (n-0, cyan line).
The retrapping current $I_{r+}$ ($I_{r-}$) and switching current $I_{c-}$ ($I_{c+}$) are defined
during the positive-to-negative (negative-to-positive) current sweep.
$I_{r+}$ ($\left|I_{r-}\right|$) is nearly the same as $I_{c+}$ ($\left|I_{c-}\right|$),
suggesting negligible capacitance in the JJ \cite{JDE_MisakiKou_PRB2021,GeneralJDE_PRX2022,Nb3Cl8FFSDE_WH_Nature2022}.
Unlike reciprocal transport ($I_{c+} = \left|I_{c-}\right|$ and $I_{r+} = \left|I_{r-}\right|$),
the difference between $I_{c+}$ ($I_{r+}$) and $\left|I_{c-}\right|$ ($\left|I_{r-}\right|$)
indicates the presence of JDE in this JJ.

The superconducting half-wave rectification is observed at $B_{z}$ = 8.4 mT and 10 mK in Fig. 1i, j.
The `on'/`off' (superconducting/normal) states are switched by alternating current bias ($I_{b} = \pm$400 nA).
Many cycles of these alternating measurements are conducted over a measurement time exceeding 1.5 hours,
demonstrating the high stability of this rectification device, which is crucial for its potential applications.
Intriguingly, the estimated switching power ($I_{b}^{2}R_{N}$) reaches the picowatt level
(6.4 pW for device S2 with $R_{N} = 40$ $\Omega$, 0.56 pW for device S1 in Supplementary Fig. 1),
which is four and eight orders of magnitude smaller than the field-free Josephson diode \cite{Nb3Cl8FFSDE_WH_Nature2022}
and another bulk superconudcting diode \cite{NbTaVSDE_AndoF_nature2020}, respectively.
This power is also close to or even lower than the power of nanowire/nanoflake SDE systems \cite{InAsSQUID_arXiv2023,InSbFlakeJDE_NanoL2022,InSbSnJDE_arXiv2022,InSbnanowire_arXiv2022}
and thin film SDE systems \cite{ThreeGrapheneFFSDE_LinJXZ_NP2022,TriodeGraphene_NanoL2023,JDEintwistBilayerGra_NC2023}.
The ultra-low switching power in Ta$_{2}$Pd$_{3}$Te$_{5}$ Josephson diodes may be attributed to fewer edge supercurrent channels, making it a potential candidate for applications.
Nonetheless, further technological advancements are required to optimize the efficiency, magnetic field conditions, and power consumption not only in this Josephson diode, but also in other superconducting diodes.

To further study the origin of JDE,
we measure the magnetic flux dependence of the JDE in the Ta$_{2}$Pd$_{3}$Te$_{5}$ JJ.
In Fig. 2a, $dV/dI$ as a function of $B_{z}$ and $I_{b}$
displays a SQUID pattern for device S1, suggesting the supercurrent interference of two channels in this JJ.
The sawtooth-shaped $I_{c}$ pattern shows up,
usually indicating the involvement of higher harmonic components due to non-negligible transmission channels \cite{BismuthHOTIsquid_NC2017},
consistent with findings from the Supplementary Note 1.
The position-dependent supercurrent density distribution (Fig. 2b) extracted from the SQUID pattern
\cite{EdgeSQUID_np2014} and the estimated enclosed area formed by two interference supercurrents
(see Supplementary Note 3 and 9) illustrate the supercurrent mainly originates from edge states of the JJ,
confirming the reported edge state in Ta$_{2}$Pd$_{3}$Te$_{5}$
\cite{Ta2Pd3Te5STM_WangXG_PRB2021,Ta2Pd3Te5LL_arXiv2022,Ta2Pd3Te5ExcitionEdge_arXiv2023}.
In Fig. 2c, the $B_{z}$-dependent $I_{c+}$ and $\left|I_{c-}\right|$ extracted
from Fig. 2a shows the asymmetric Josephson effect.
Several features, such as the large deviation of $I_{c}$ from zero at half flux quantum in Fig. 2c
and JDE in Fig. 2d, indicate the asymmetric edge supercurrent channels in this JJ, as detailed in methods.
Moreover, $\Delta I_{c} = I_{c+} - \left|I_{c-}\right|$ oscillates with $B_{z}$ (orange curve in Fig. 2d),
which can be called interfering JDE and is approximately described by the two-JJs model (black line in Fig. 2d,
and see the detail in methods).
This type of JDE has been theoretically predicted in Weyl semimetals with broken inversion symmetry
and asymmetric helical edge states \cite{AsymmetricJDE_ChenCZ_PRB2018}.
Furthermore, the oscillating Josephson diode efficiency
(cyan line in Fig. 2d) shows a maximal efficiency $\eta \approx$ 45 \% at 6.4 mT.
It is intriguing that $\eta$ can reach 10 \% at $B_{z}$ = -0.5 mT.
Also, the required field can be significantly reduced when the enclosed area of the SQUID loop is increased,
which could be easily realized in the experimental setup \cite{TiSQUIDSDE_PaolucciF_APL2023,InAsSQUID_arXiv2023}.

\begin{figure*}[!thb]
\begin{center}
\includegraphics[width=7in]{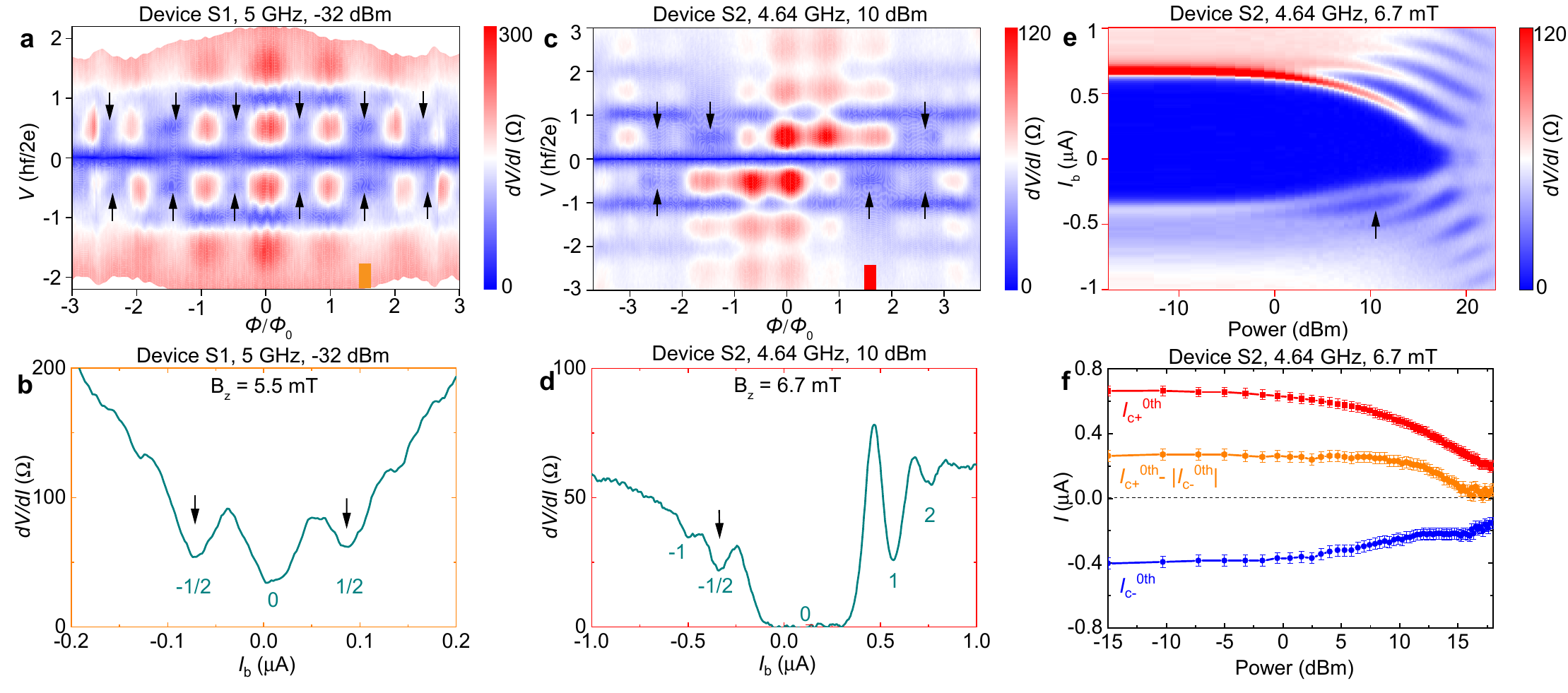}
\end{center}
\caption{\label{Fig3} \textbf{Fractional Shapiro steps under microwave in both Ta$_{2}$Pd$_{3}$Te$_{5}$ devices.}
$\mathbf{a}$, Differential resistance ($dV/dI$) as a function of voltage characteristics
and flux quantum at microwave power = -32 dBm, $f$ = 5 GHz and 30 mK for device S1.
The $\pm1/2$th Shapiro steps are marked by black arrows.
$\mathbf{b}$, $dV/dI$ versus $I_{b}$ at $B_{z}$ = 5.5 mT (3/2 $\phi_{0}$).
Valleys characterized by half-integer steps are observed clearly.
$\mathbf{c}$, Flux quantum and voltage dependence of $dV/dI$ at microwave power = 10 dBm, $f$ = 4.64 GHz and 10 mK for device S2.
$\mathbf{d}$, Valleys characterized by the $-1/2$th Shapiro step at 6.7 mT (3/2 $\phi_{0}$).
$\mathbf{e}$, Microwave power dependence of $dV/dI$ at 4.64 GHz and $B_{z}$ = 6.7 mT in device S2.
The $- 1/2$th Shapiro step is marked by black arrows.
$\mathbf{f}$, Microwave-power-dependent non-reciprocal critical current of the 0th Shapiro steps,
which is extracted from $\mathbf{e}$.
The switching currents here can be determined by peaks in the $dV/dI$ - $I_b$ curve, as shown in Supplementary Fig. 8c,
and the error bars primarily originate from the broadening of these peaks.
}
\end{figure*}

Having demonstrated the JDE in Ta$_{2}$Pd$_{3}$Te$_{5}$ edge interferometer,
we now show how to further enhance its efficiency.
Interestingly, compared with device S1,
the SQUID pattern of device S2 in Fig. 2f reveals two main periods,
leading to a more pronounced asymmetry in the pattern (Fig. 2h, i)
and an increased efficiency with $\eta_{max}$ = 73 \% at 8.4 mT.
The enhanced JDE can be explained by three JJs in parallel.
The formation of the middle edge state in the JJ may stem from the ladder-like structure induced by imperfect exfoliation (as indicated by the green arrow in the inset of Fig. 2g and supplementary Fig. 3d),
which is a common occurrence during the exfoliation process of 2D flakes \cite{CrackinmonolayerWTe2_SA2019} .
These three supercurrent channels form two sets of interference patterns
(corresponding red and green dashed loops in the inset of Fig. 2g),
leading to the asymmetric SQUID. The $\Delta I_{c}$ can also be approximately fitted by the three-JJs model (black line in Fig. 2i, and see the detail in methods).
The optimization of $\eta$ in device S2 (Fig. 2j), compared to device S1 in Fig. 2e, seems to be achieved by increasing the number of JJs in parallel, which is in line with the theoretically predicted method \cite{FDEinSI_Constanti_PRL2022}.

Here, we examine the main mechanisms to determine which one best fits our data.
(1) Magnetism mechanism \cite{NbVCoVTaFFSDE_NN2022} is initially excluded due to the absence of magnetism in the Ta$_{2}$Pd$_{3}$Te$_{5}$ JJ.
(2) Rashba-spin orbital coupling (SOC) mechanism is excluded as it typically requires in-plane magnetic fields,
and Ising-SOC is also considered implausible because it is usually found in 2D superconductors \cite{2DSC_SatioY_NRM2017}.
(3) JDE/SDE caused by finite-momentum Cooper pairing generally also needs
in-plane magnetic fields \cite{SDEFMS_YuanNFQ_PNAS2022,NiTe2JDE_PalB_NP2022}.
The absence of obvious enhancement of the upper critical field under in-plane magnetic fields or the approximately butterfly interference pattern in our JJs \cite{NiTe2JDE_PalB_NP2022} (Supplementary Fig. 5) leads to the exclusion of finite-momentum Cooper pairing as a mechanism.
(4) Nonlinear capacitance \cite{JDE_MisakiKou_PRB2021,JDEmageticAtton_TrahmsM_nature2023} is absent in this system
due to the symmetric superconducting electrodes and the lack of obvious hysteresis
between switching current and retrapping current (Fig. 1h).
(5) Self-inductance is also found to be too small to be considered (see Supplementary Note 8).
Therefore, the main origin of $B_{z}$-induced JDE in the Ta$_{2}$Pd$_{3}$Te$_{5}$ edge interferometer is
attributed to asymmetric edge supercurrents, as detailed in methods.

\vspace{3ex}
\noindent\textbf{Fractional Shapiro steps}

As discussed above, besides asymmetric edge supercurrents and time-reversal symmetry breaking,
the remaining higher harmonic in the CPR is further studied to explain the JDE in our JJs.
The Shapiro resonances have been widely used to reveal not only
the exotic symmetry of Cooper pairing \cite{4piPeriodJJ_NC2016}   
but also the higher harmonic or non-sinusoidal CPR \cite{FDEinSI_Constanti_PRL2022,FST_Chauvin_PRL2006}.
The Shapiro steps appear in the $I$ - $V$ curve at $V = V_{n} \equiv nhf/2e$
($n$ is an integer) when the JJ is irradiated with the microwave.
If the $n$th harmonic contributes to the CPR, $m/n$th Shapiro steps ($m$ is an integer) will be present \cite{FDEinSI_Constanti_PRL2022,FST_Chauvin_PRL2006}.
In Fig. 3a, fractional Shapiro steps emerge at a microwave frequency of $f$ = 5 GHz,
indicating the significant transparency of the JJ (Supplementary Note 1).
The $\pm1/2$th Shapiro steps at half flux quantum $(n+1/2)\phi_{0}$ are marked by black arrows,
which is also observed in other works \cite{InAsAlSQUIDjde_arXiv2023,2DgasSQUIDdiode_arXiv2023}.
The characteristic $dV/dI$ valleys formed by half-integer steps can be clearly observed
at marked positions in Fig. 3b at $B_{z}$ = 5.5 mT (corresponding to 3/2 $\phi_{0}$).
In addition, half-integer Shapiro steps are also observed in device S2 with flux-dependent measurements
($f$ = 4.64 GHz in Fig. 3c, d or 5.02 GHz in Supplementary Fig. 7)
and power-dependent measurements at $3\phi_{0}/2$ (Fig. 3e).
These experiments indicate the existence of the second harmonic in the CPR,
which is also supported by the rapid suppression observed in temperature-dependent $\Delta I_{c}$
compared with $I_{c}$ in device S1 (see Fig. 2e),
because higher harmonics are quickly suppressed when temperature approaches $T_{c}$ \cite{Al2DEG_BaumgartnerC_NN2022}.
The rapid suppression of $\Delta I_{c}$ in device S2 (Fig. 2j) seems less pronounced,
possibly because its three-JJs model enhances the diode effect, therefore relatively weakening such suppression.
In addition to their applications in Josephson diodes, the JJs with higher harmonics also have potential applications in 0 - $\pi$ qubits, particularly if the first harmonic in the CPR can be effectively eliminated \cite{Qubit_PRL2020,SCQubit_PRXQ2022,SCQubit2_PRXQ2022}.

Furthermore, some features in Fig. 3 require further explanation.
First, the disappearance of half-integer Shapiro steps at zero field (Fig. 3a, c) may be attributed to the relatively small contribution of the second harmonic in the CPR and the low microwave frequency \cite{FractionalShapiro_PRB2020}.
In device S1, a higher microwave frequency is utilized at zero field, and a small signal of half-integer steps appears in Supplementary Fig. 6d-f. No higher frequency is applied to detect half-integer steps at 0 T, considering the limitations of our microwave generator.
Second, the integer Shapiro steps at 5.5 mT (3/2 $\phi_{0}$) appear weaker in Fig. 3b (or Supplementary Fig. 6c) compared to those at 0 T in Supplementary Fig. 6b. This is due to the application of a magnetic field at half the flux quantum,
causing a reduction in the sharpness of the integer steps because of a decrease in the first harmonic of the CPR.
Subsequently, the second harmonic or fractional Shapiro steps become visible at half the flux quantum (Fig. 3a), in agreement with theoretical predictions \cite{FDEinSI_Constanti_PRL2022,JDESQUID_Fominov_PRB2022} and experimental results \cite{InAsAlSQUIDjde_arXiv2023,2DgasSQUIDdiode_arXiv2023}.
Third, in device S2 (Fig. 3c), half-integer steps vanish at $\pm\phi_{0}/2$
due to the superposition of multiple SQUID patterns,
which is slightly different from the single SQUID pattern observed in device S1.
In the case of multiple SQUID patterns, the position of destructive interference for the first harmonic differs from that of a single SQUID pattern, leading to a notable contribution from the first harmonic and the absence of half-integer steps at $\pm\phi_{0}/2$ (see Supplementary Note 6).

\begin{figure*}[!thb]
\begin{center}
\includegraphics[width=7in]{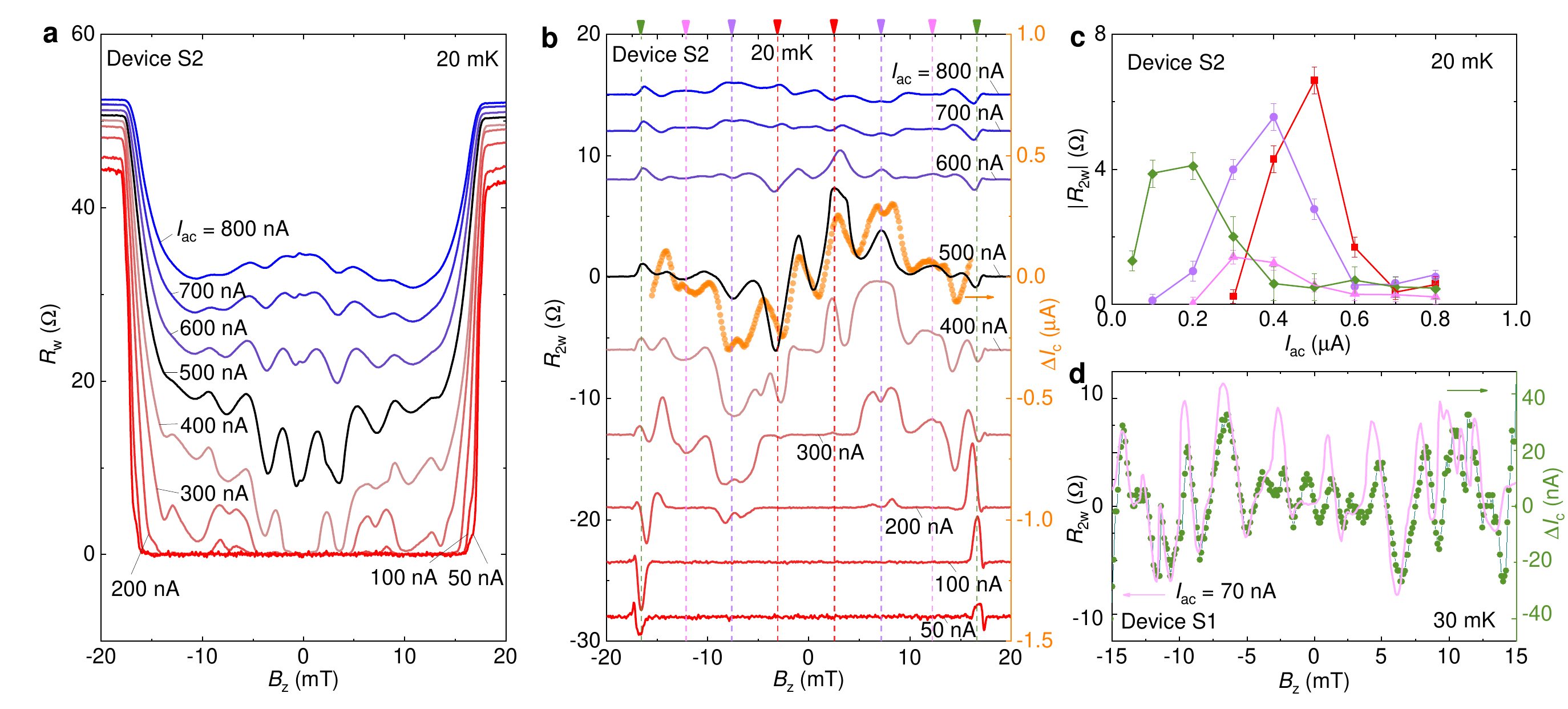}
\end{center}
\caption{\label{Fig4} \textbf{Antisymmetric second harmonic transport of Ta$_{2}$Pd$_{3}$Te$_{5}$ JJs.}
$\mathbf{a}$, $R_{w}$ versus $B_{z}$ at different currents.
$\mathbf{b}$, $B_{z}$-dependent $R_{2w}$ at various currents.
Typical pairs of antisymmetric peaks ($R_{2w}$) are marked by dashed lines in matching colors.
$R_{2w}$ at 500 nA (black line) shows a similar behavior to $\Delta I_{c}$ (orange line).
The green dashed line indicates the regions near the critical fields,
while the other dashed line marks the areas near the oscillating peaks or valleys.
$\mathbf{c}$, Typical $\left| R_{2w}\right|$ as a function of current at 20 mK for different oscillating peaks.
The average $\left| R_{2w}\right|$ values are obtained using pairs of antisymmetric peaks near the dashed lines in $\mathbf{b}$.
The error bars primarily originate from discrepancies between average $\left| R_{2w}\right|$ and the observed values.
$\mathbf{d}$, Antisymmetric $R_{2w}$ and $\Delta I_{c}$ as a function of $B_{z}$ in device S1.
The $R_{2w}$ is measured at $I_{ac}$ = 70 nA.
    }
\label{Fig4}
\end{figure*}

The JDE is also studied as a function of microwave power.
In Fig. 3f, switching current $I_{c+}^{0th}$ and $I_{c-}^{0th}$ of the
0th Shapiro step are extracted from the Fig. 3e.
The quantity $\Delta I_{c}^{0th} = I_{c+}^{0th} - \left|I_{c-}^{0th}\right|$
(orange curve, see the detail in Supplementary Fig. 8)
displays little variation and gradually decreases with increasing microwave power.
This stable observation of the JDE under microwave irradiation represents one of the initial examples of such behavior and is a crucial feature in quantum information applications, where microwaves are commonly utilized \cite{SQC2_YuY_science2002}.

\vspace{3ex}
\noindent\textbf{Antisymmetric second harmonic transport}

As another type of non-reciprocal behavior, the second harmonic resistance ($R_{2w}$) is usually measured to characterize magnetochiral anisotropy (MCA) \cite{MoS2NCT_WakatsukiR_SA2017,NbTaVSDE_AndoF_nature2020,Nb3Cl8FFSDE_WH_Nature2022},
whose strength is often defined by the coefficient $\gamma = \left|\frac{2R_{2w}}{R_{w}BI}\right|$.
MCA typically exhibits antisymmetric behavior in $R_{2w}$ with respect to magnetic fields
and is generally used to characterize noncentrosymmetric or chiral features.
In normal conductors, MCA is usually a minor effect, generally due to the weak SOC. 
However, a significant enhancement of MCA is observed in noncentrosymmetric superconductors,
potentially offering additional insights into factors such as disparities between the Fermi energy and superconducting gap, the Rashba parameter, and spin-orbit splitting, among other factors \cite{MoS2NCT_WakatsukiR_SA2017,NTin2DSC_PRB2018}.
Moveover, the antisymmetric oscillations in $B$-dependent $R_{2w}$ have been observed in an ionic liquid-gated chiral nanotube exhibiting the Little-Parks effect \cite{NanotubeLPR2w_NC2017}, indicating its characteristic chiral symmetry.

Therefore, the second harmonic resistance in our supercurrent interferometer is measured to clarify its asymmetric features.
In Fig. 4a, b, the first harmonic resistance ($R_{w}$) and $R_{2w}$ of device S2 are measured
using lock-in techniques.
Below $I_{ac}$ = 100 nA, a typical pair of antisymmetric peaks in $R_{2w}$ is observed at critical fields
(near the green dashed line in Fig. 4b).
This is consistent with the previously reported antisymmetric peaks of $R_{2w}$ with respect to $B$ in noncentrosymmetric superconductors \cite{MoS2NCT_WakatsukiR_SA2017,NbTaVSDE_AndoF_nature2020}, suggesting the existence of asymmetry in our JJ.
With increasing $I_{ac}$, the antisymmetric oscillations in $R_{2w}$ becomes more pronounced
due to the quantum interference of asymmetric edge supercurrents,
similar to the behavior observed in chiral nanotubes \cite{NanotubeLPR2w_NC2017}.
At around 500 nA, the oscillating $R_{2w}$ (black line in Fig. 4b) is analogous to
the $\Delta I_{c}$ (orange line), and the similar behavior of device S1 can be seen in Fig. 4d
(see the detail in Supplementary Note 7).
This similarity suggests the antisymmetric oscillations in $R_{2w}$ may be related to the polarity of JDE.
At higher $I_{ac}$, the superconductivity is substantially suppressed and self-heating effects
may begin to influence the second harmonic resistance, which is not further analyzed.

In addition, supercurrent interference-induced $R_{2w}$ in our JJs appears more complex.
In Fig. 4c, the current-dependent $R_{2w}$ (red, magenta and violet lines) extracted from the same oscillating peaks below critical fields does not intersect the origin, suggesting deviations from the $R_{2w}$ = $\frac{1}{2}\gamma R_{w}IB$ formula generally observed in other systems \cite{SrTiO3GateNCT_ItahashiYM_SA2020,SrTiO3LaTiO3nonreciprocal_NC2019,BiSbTe3_LeggHF_NN2022,NbSe2NCT_ZhangEZ_NC2020}.
Therefore, although the antisymmetric oscillations in $R_{2w}$ at large $I_{ac}$ display partial antisymmetric features of MCA,
the 2$R_{2w}/R_{w}IB$ calculated using the typical formula below the critical field may not accurately
describe $\gamma$ or MCA. The MCA in interference systems appears more complex and requires further investigation.

\vspace{3ex}
\noindent\textbf{Discussion}

The Ta$_{2}$Pd$_{3}$Te$_{5}$ edge interferometer
reveals a significant JDE under small out-of-plane magnetic fields.
The JDE is accompanied by the enhanced antisymmetric second harmonic transport, which deserves further research.
Moreover, the presence of asymmetric Josephson critical current alongside fractional Shapiro steps
demonstrates the importance of the higher harmonic content in the JDE.
It is noted that power consumption requires attention if SDE/JDE is to be applied at low temperatures in the future,
and further optimization is necessary not only regarding mechanism, but also in terms of device size.
The Josephson diode efficiency can be further enhanced by concatenating interferometer loops.
These findings offer a promising approach to exploring JDE with significant diode efficiency,
ultra-low switching power at small magnetic fields and stable JDE under microwave
irradiation, rendering them potentially valuable for practical applications.

\vspace{3ex}

\noindent\textbf{Methods}

\noindent\textbf{Device fabrication.} Single crystals Ta$_{2}$Pd$_{3}$Te$_{5}$ were prepared by
the self-flux method \cite{Ta2Pd3Te5QSH_GZP_PRB21}. Ta$_{2}$Pd$_{3}$Te$_{5}$ thin films were obtained
through mechanical exfoliating bulk samples onto Si substrates with the 280-nm-thick SiO$_{2}$ on top,
and then coated with PMMA in a glove box at 4000 rpm for 60 s, followed by annealing at 100 $^{\circ}$C for 120 s.
Multi-terminal electrical contacts were patterned after electron-beam exposure and subsequent development.
Ti/Al (5 nm/60 nm) electrodes were deposited after Ar etching in order to remove the oxidized layer.
After the lift-off step, the device was coated with hexagonal boron nitride (hBN) to prevent oxidation.
The entire fabrication process took place in a nitrogen atmosphere glove box.
The rectangular film, with a long side along the $b$-axis of the crystal,
was easily obtained due to its quasi-1D nature \cite{Ta2Pd3Te5LL_arXiv2022}.
The dimensions of device S2 (device S1) are as follows: thickness 37 nm (16 nm),
width 1.9 $\mu$m (1.3 $\mu$m), and length 245 nm (400 nm), respectively.

\vspace{3ex}

\noindent\textbf{Transport measurements.} The electrical transport measurements were carried out
in cryostats (Oxford instruments dilution refrigerator).
The dc current bias measurements were applied using a Keithley 2612 current source.
Both the first- ($R_{w}$) and second-harmonic ($R_{2w}$) resistance
were measured through a standard low-frequency (7-11 Hz) lock-in technique (LI5640, NF Corporation).
The phase of the first- and second-harmonic signal was set to be 0$^{\circ}$ and 90$^{\circ}$, respectively.
The current of magnet was applied by a Keithley 2400 current source
in order to control the magnetic field accurately.
Before the measurement, annealing process with 300 $^{\circ}$C for $\sim$10 minutes
was performed to obtain good contacts in the Ta$_{2}$Pd$_{3}$Te$_{5}$ JJs.
$I-V$ curves were obtained though the numerical integration process,
and the origin data were all $dV/dI-I_{b}$ curves.

\vspace{3ex}

\noindent\textbf{Transport measurements under microwave radiation.}
The electrical transport measurements under microwave radiation were carried out
using PSG-A Series Signal generator (Agilent E8254A).
The main microwave frequency between 4.64 and 5.02 GHz was applied due to
the large absorption effect of microwave for devices S1 and S2.
The dissipation of microwave circuit existed during the measurement of device S2,
while another microwave circuit for device S1 was good.
So the applied power of microwave was a little large
but without inducing obvious thermal effect.

\vspace{3ex}

\noindent\textbf{Band structure calculations.}
The first-principles calculations were carried out based on the density functional theory
(DFT) with the projector augmented wave (PAW) method, 
as implemented in the Vienna \emph{ab initio} simulation package (VASP) \cite{Ta2Pd3Te5TI_GuoZP_npjqm2022}.
The generalized gradient approximation (GGA) in the form of Pardew-Burke-Ernzerhof (PBE) function 
was employed for the exchange-correlation potential.
The kinetic energy cutoff for plane wave expansion was set to 400 eV,
and a 1$\times$9$\times$1 $\textbf{k}$-mesh was adopted for the Brillouin zone sampling in the self-consistent process.
The thicknesses of the vacuum layer along $x$ and $z$ axis were set to \textgreater ~20 \AA.

\vspace{3ex}

\noindent\textbf{Simulations of JDE.}
A minimal model with $n$ JJs in parallel is used to explain the JDE \cite{FDEinSI_Constanti_PRL2022},
where only the first and second harmonic contributions in the CPR are considered.
The supercurrent is approximatively described as $I (\varphi,\phi/\phi_{0}) = \sum_{1}^n I_{n}sin(\varphi + 2\pi\zeta_{n}\phi/\phi_{0}) + \alpha_{n}I_{n}sin(2\varphi + 4\pi\zeta_{n}\phi/\phi_{0})$,
where $\zeta_{n}$ is the modified ratio and $\zeta_{1}$ = 0, $\zeta_{2}$ = 1
($\zeta_{1}$ = 0, $\zeta_{2}$ = 1, $\zeta_{3}$ = 0.23) for the simulation of our two (three) JJs in parallel.
$\zeta_{3}$ represents the enclosed-area ratio of two corresponding SQUID patterns in device S2.
Preliminary switching currents for different channels are estimated in both devices
without considering the second harmonic contribution, in order to simulate CPR and JDE.
For device S1, the parameters used for simulating $\Delta I_{c}$ are: $I_{1}$ = 75 nA, $I_{2}$ = 15 nA,
$\alpha_{1}$ = 0.3 and $\alpha_{2}$ = 0.5, while for device S2, the parameters are: $I_{1}$ = 330 nA,
$I_{2}$ = 120 nA, $I_{3}$ = 158 nA, $\alpha_{1}$ = 0.6, $\alpha_{2}$ = 0.1 and $\alpha_{3}$ = 0.2.
The simulated $\phi/\phi_{0}$-dependent $\Delta I_{c}$ can approximatively explain the experimental data in both devices,
as shown by black lines in Fig. 2d, i.
The slight discrepancy between the simulation and experiment may arise from the omission of higher-order harmonics in the CPR, a reduction in critical current and a potential variation of $\alpha$ with increasing magnetic field, or other contributing factors.

\vspace{3ex}

\noindent\textbf{Asymmetric edge supercurrents.} We will discuss edge supercurrents,
their asymmetry, and the origin of such asymmetry in our cases, using device S1 as an example.
Firstly, the SQUID-like pattern observed in Fig. 2a exhibits several interference features of two edge supercurrent channels, as supported by previous reports \cite{Ta2Pd3Te5STM_WangXG_PRB2021,Ta2Pd3Te5LL_arXiv2022,Ta2Pd3Te5ExcitionEdge_arXiv2023}.
In this pattern, both the width and amplitude of the center lobe are approximately equal to those of the other lobes.
This is obviously contrast to a typical Fraunhofer pattern induced by the bulk transport \cite{EdgeSQUID_np2014,EdgeSC_NN2015},
where the width of the center lobe is twice as large as that of the other lobes, while the amplitude of the center lobe is usually significantly larger than that of the other lobes.

Secondly, there are several features supporting asymmetric edge supercurrents in our devices:
(1) Without considering the higher harmonics of CPR,
we calculated two edge supercurrent channels ($I_{1}$ and $I_{2}$) from the SQUID pattern in Supplementary Note 3, and in device S1, the ratio between the supercurrents $I_{1}$:$I_{2}$ is approximately 3.5:1 ($I_{1}$:$I_{2}$:$I_{3}$ $\sim$ 6:1:2 for device S2);
(2) After considering the second harmonic, from a simple simulation using a two-JJs model in Supplementary Fig. 2e, the deviation of $I_{c}$ from zero at half flux quantum alone cannot support asymmetric interference channels. However, the coexistence of this deviation and the JDE can support asymmetric supercurrents, as shown in Supplementary Fig. 2d, g, h, and device S1 satisfies this coexistence condition;
(3) The position-dependent supercurrent density distribution shows a significantly asymmetric supercurrent distribution. However, it should be noted that such asymmetric supercurrent density distribution can also exist in a symmetric system \cite{EdgeSQUID_np2014,EdgeSC_NN2015} due to the relatively imperfect experimental data.

Finally, the origin of asymmetric edge supercurrents is further discussed:
(1) The DFT calculations in Fig. 1g support the existence of asymmetric edge states,
which contribute to asymmetric edge supercurrents and JDE \cite{AsymmetricJDE_ChenCZ_PRB2018};
(2) Different transmission coefficients \cite{FDEinSI_Constanti_PRL2022} of two edge JJs may also lead to asymmetric edge supercurrents,
and can be caused by the different edge disorders, different coupling with the Al contacts, and other factors.
Such conditions were not intentionally controlled in our devices.

\vspace{3ex}

\noindent\textbf{Discussion of switching power.} From a practical perspective,
the power consumption of devices should be carefully considered at low temperatures.
Although the current cooling power of commercial dilution refrigerators generally reaches the microwatt level
at 10 mK \cite{Dilutionrefrigerator_Cryogenics2022},
a Joule excitation power higher than a picowatt (or nanowatt) may raise the temperature of the device or sensor above 10 mK (or 100 mK) when the mixing chamber is kept at 10 mK \cite{RuO2downto5mK_Cryogenics2021,ThermoR_AIPCP2002}.
The self-heating effect is sometimes noticeable and can also be applied.
For example, a microheater with a power output of a few picowatts can generate a temperature gradient as large as several millikelvin per micron, and can be further used to conduct Seebeck or Nernst effect measurements in 2D devices at millikelvin base temperatures \cite{NernstinWTe2_arXiv23}.
Therefore, ultra-low power consumption is essential not only for basic measurements but also for applications in large-scale superconducting quantum circuits, which need ultra-low temperatures to enhance the precision of quantum state measurements.

The edge-induced JDE in Ta$_{2}$Pd$_{3}$Te$_{5}$ may offer distinct advantages in terms of power consumption.
The edges of the topological system exhibit a low number of conducting channels compared to bulk states transport.
This characteristic can result in a lower critical current and reduced power consumption \cite{EdgeSQUID_np2014,EdgeSC_NN2015}.

\vspace{3ex}

\noindent\textbf{Data availability}

\noindent All relevant data are available from the authors. The data can also be found at the following link (http://dx.doi.org/10.6084/m9.figshare.26539759).

\vspace{3ex}

\noindent\textbf{Code availability}

\noindent DFT calculations can be reproduced using standard VASP packages. Simulations of JDE are fully described.
The codes used in this study are available from the authors upon request.

\vspace{3ex}

\noindent\textbf{Acknowledgments}

We thank Jin-Guang Cheng for deep discussions. This work was supported by the Beijing Natural Science Foundation (Grant No. JQ23022),
the National Natural Science Foundation of China
(Grant Nos. 92065203, 12174430, U2032204, 11974395, 12188101, U22A6005 and 12404154),
the Strategic Priority Research Program of Chinese Academy of Sciences (Grant No. XDB33000000 and XDB33030000),
the Beijing Nova Program (Grant No. Z211100002121144),
the Synergetic Extreme Condition User Facility (SECUF),
the Informatization Plan of Chinese Academy of Sciences (CAS-WX2021SF-0102),
the Ministry of Science and Technology of China (Grants No. 2022YFA1403800),
the Financial Support from Innovation Program for Quantum Science and Technology (Grant No. 2021ZD0302500),
the China Postdoctoral Science Foundation (Grant No. 2021M703462 and 2021TQ0356)
and the Center for Materials Genome.
A portion of this work was carried out at the Synergetic Extreme Condition User Facility (SECUF).

\vspace{3ex}

\vspace{3ex}

\noindent\textbf{Author contributions}

\noindent J. S. and Y.P.L. conceived and designed the experiment.
Y.P.L. and Y.H. fabricated devices with the help of A.Q.W., X.C.G., X.F.S. and Z.K.S..
Y.P.L. and Y.H. performed the transport measurements,
supervised by Z.W.D., Z.Z.L., T.Q., G.T.L, F.M.Q., Z.A.X., L.L. and J. S..
D.Y.Y. and Y.G.S. synthesized bulk Ta$_{2}$Pd$_{3}$Te$_{5}$ crystals.
H.H.S. and Z.J.W. calculated the band structure.
J.K. and J.P.H. provided some supports on theoretical modeling.
Y.P.L. and J.S. wrote the manuscript,
and all authors contributed to the discussion of results and improvement of the manuscript.

\vspace{3ex}

\noindent\textbf{Competing interests}
\noindent The authors declare no competing interests.


\vspace{3ex}

\noindent\textbf{References}
\begin{thebibliography}{10}
\expandafter\ifx\csname url\endcsname\relax
  \def\url#1{\texttt{#1}}\fi
\expandafter\ifx\csname urlprefix\endcsname\relax\def\urlprefix{URL }\fi
\providecommand{\bibinfo}[2]{#2}
\providecommand{\eprint}[2][]{\url{#2}}

\bibitem{PNjunction_shockley_BSTJ1949}
\bibinfo{author}{{Shockley}, W.}
\newblock \bibinfo{title}{The theory of p-n junctions in semiconductors and p-n
  junction transistors}.
\newblock \emph{\bibinfo{journal}{Bell Syst. Tech. J.}}
  \textbf{\bibinfo{volume}{28}}, \bibinfo{pages}{435--489}
  (\bibinfo{year}{1949}).

\bibitem{SQC2_YuY_science2002}
\bibinfo{author}{{Yu}, Y.}, \bibinfo{author}{{Han}, S.},
  \bibinfo{author}{{Chu}, X.}, \bibinfo{author}{{Chu}, S.-I.} \&
  \bibinfo{author}{{Wang}, Z.}
\newblock \bibinfo{title}{{Coherent temporal oscillations of macroscopic
  quantum states in a {J}osephson junction}}.
\newblock \emph{\bibinfo{journal}{Science}} \textbf{\bibinfo{volume}{296}},
  \bibinfo{pages}{889--892} (\bibinfo{year}{2002}).

\bibitem{SQC3_IoffeLB_Nature1999}
\bibinfo{author}{{Ioffe}, L.~B.}, \bibinfo{author}{{Geshkenbein}, V.~B.},
  \bibinfo{author}{{Feigel'Man}, M.~V.}, \bibinfo{author}{{Fauch{\`e}re},
  A.~L.} \& \bibinfo{author}{{Blatter}, G.}
\newblock \bibinfo{title}{{Environmentally decoupled sds-wave {J}osephson
  junctions for quantum computing}}.
\newblock \emph{\bibinfo{journal}{Nature}} \textbf{\bibinfo{volume}{398}},
  \bibinfo{pages}{679--681} (\bibinfo{year}{1999}).

\bibitem{JDE_HJP_PRL2007}
\bibinfo{author}{{Hu}, J.}, \bibinfo{author}{{Wu}, C.} \&
  \bibinfo{author}{{Dai}, X.}
\newblock \bibinfo{title}{{Proposed design of a {J}osephson diode}}.
\newblock \emph{\bibinfo{journal}{Phys. Rev. Lett.}}
  \textbf{\bibinfo{volume}{99}}, \bibinfo{pages}{067004}
  (\bibinfo{year}{2007}).

\bibitem{AsymmetricJDE_ChenCZ_PRB2018}
\bibinfo{author}{{Chen}, C.-Z.} \emph{et~al.}
\newblock \bibinfo{title}{{Asymmetric {J}osephson effect in inversion symmetry
  breaking topological materials}}.
\newblock \emph{\bibinfo{journal}{Phys. Rev. B}} \textbf{\bibinfo{volume}{98}},
  \bibinfo{pages}{075430} (\bibinfo{year}{2018}).

\bibitem{JDE_MisakiKou_PRB2021}
\bibinfo{author}{{Misaki}, K.} \& \bibinfo{author}{{Nagaosa}, N.}
\newblock \bibinfo{title}{{Theory of the nonreciprocal Josephson effect}}.
\newblock \emph{\bibinfo{journal}{Phys. Rev. B}}
  \textbf{\bibinfo{volume}{103}}, \bibinfo{pages}{245302}
  (\bibinfo{year}{2021}).

\bibitem{ISDE_Daido_PRL2022}
\bibinfo{author}{{Daido}, A.}, \bibinfo{author}{{Ikeda}, Y.} \&
  \bibinfo{author}{{Yanase}, Y.}
\newblock \bibinfo{title}{{Intrinsic superconducting diode effect}}.
\newblock \emph{\bibinfo{journal}{Phys. Rev. Lett.}}
  \textbf{\bibinfo{volume}{128}}, \bibinfo{pages}{037001}
  (\bibinfo{year}{2022}).

\bibitem{SDEFMS_YuanNFQ_PNAS2022}
\bibinfo{author}{{Yuan}, N. F.~Q.} \& \bibinfo{author}{{Fu}, L.}
\newblock \bibinfo{title}{{Supercurrent diode effect and finite-momentum
  superconductors}}.
\newblock \emph{\bibinfo{journal}{Proc. Natl Acad. Sci. USA}}
  \textbf{\bibinfo{volume}{119}}, \bibinfo{pages}{e2119548119}
  (\bibinfo{year}{2022}).

\bibitem{FDEinSI_Constanti_PRL2022}
\bibinfo{author}{{Souto}, R.~S.}, \bibinfo{author}{{Leijnse}, M.} \&
  \bibinfo{author}{{Schrade}, C.}
\newblock \bibinfo{title}{{Josephson diode effect in supercurrent
  interferometers}}.
\newblock \emph{\bibinfo{journal}{Phys. Rev. Lett.}}
  \textbf{\bibinfo{volume}{129}}, \bibinfo{pages}{267702}
  (\bibinfo{year}{2022}).

\bibitem{JDE_FuLiang_SA2022}
\bibinfo{author}{{Davydova}, M.}, \bibinfo{author}{{Prembabu}, S.} \&
  \bibinfo{author}{{Fu}, L.}
\newblock \bibinfo{title}{{Universal {J}osephson diode effect}}.
\newblock \emph{\bibinfo{journal}{Sci. Adv.}} \textbf{\bibinfo{volume}{8}},
  \bibinfo{pages}{eabo0309} (\bibinfo{year}{2022}).

\bibitem{GeneralJDE_PRX2022}
\bibinfo{author}{{Zhang}, Y.}, \bibinfo{author}{{Gu}, Y.},
  \bibinfo{author}{{Li}, P.}, \bibinfo{author}{{Hu}, J.} \&
  \bibinfo{author}{{Jiang}, K.}
\newblock \bibinfo{title}{{General theory of Josephson diodes}}.
\newblock \emph{\bibinfo{journal}{Phys. Rev. X}} \textbf{\bibinfo{volume}{12}},
  \bibinfo{pages}{041013} (\bibinfo{year}{2022}).

\bibitem{SDE_HJJ_NJP2022}
\bibinfo{author}{{He}, J.~J.}, \bibinfo{author}{{Tanaka}, Y.} \&
  \bibinfo{author}{{Nagaosa}, N.}
\newblock \bibinfo{title}{{A phenomenological theory of superconductor
  diodes}}.
\newblock \emph{\bibinfo{journal}{New J.Phys.}} \textbf{\bibinfo{volume}{24}},
  \bibinfo{pages}{053014} (\bibinfo{year}{2022}).

\bibitem{NbTaVSDE_AndoF_nature2020}
\bibinfo{author}{Ando, F.} \emph{et~al.}
\newblock \bibinfo{title}{Observation of superconducting diode effect}.
\newblock \emph{\bibinfo{journal}{Nature}} \textbf{\bibinfo{volume}{584}},
  \bibinfo{pages}{373--376} (\bibinfo{year}{2020}).

\bibitem{NiTe2JDE_PalB_NP2022}
\bibinfo{author}{{Pal}, B.} \emph{et~al.}
\newblock \bibinfo{title}{{Josephson diode effect from {C}ooper pair momentum
  in a topological semimetal}}.
\newblock \emph{\bibinfo{journal}{Nat. Phys.}} \textbf{\bibinfo{volume}{18}},
  \bibinfo{pages}{1228--1233} (\bibinfo{year}{2022}).

\bibitem{ThreeGrapheneFFSDE_LinJXZ_NP2022}
\bibinfo{author}{{Lin}, J.-X.} \emph{et~al.}
\newblock \bibinfo{title}{{Zero-field superconducting diode effect in
  small-twist-angle trilayer graphene}}.
\newblock \emph{\bibinfo{journal}{Nat. Phys.}} \textbf{\bibinfo{volume}{18}},
  \bibinfo{pages}{1221--1227} (\bibinfo{year}{2022}).

\bibitem{Y3Fe5O12FFJDE_JeonKR_NM2022}
\bibinfo{author}{{Jeon}, K.-R.} \emph{et~al.}
\newblock \bibinfo{title}{{Zero-field polarity-reversible Josephson
  supercurrent diodes enabled by a proximity-magnetized Pt barrier}}.
\newblock \emph{\bibinfo{journal}{Nat. Mater.}} \textbf{\bibinfo{volume}{21}},
  \bibinfo{pages}{1008--1013} (\bibinfo{year}{2022}).

\bibitem{NbSe2SCE_BauriedlL_NC2022}
\bibinfo{author}{{Bauriedl}, L.} \emph{et~al.}
\newblock \bibinfo{title}{{Supercurrent diode effect and magnetochiral
  anisotropy in few-layer {NbSe}$_2$}}.
\newblock \emph{\bibinfo{journal}{Nat. Common.}} \textbf{\bibinfo{volume}{13}},
  \bibinfo{pages}{4266} (\bibinfo{year}{2022}).

\bibitem{JDEintwistBilayerGra_NC2023}
\bibinfo{author}{{D{\'\i}ez-M{\'e}rida}, J.} \emph{et~al.}
\newblock \bibinfo{title}{{Symmetry-broken {J}osephson junctions and
  superconducting diodes in magic-angle twisted bilayer graphene}}.
\newblock \emph{\bibinfo{journal}{Nat. Commun.}} \textbf{\bibinfo{volume}{14}},
  \bibinfo{pages}{2396} (\bibinfo{year}{2023}).

\bibitem{InSbFlakeJDE_NanoL2022}
\bibinfo{author}{{Turini}, B.} \emph{et~al.}
\newblock \bibinfo{title}{{Josephson diode effect in high-mobility {InSb}
  nanoflags}}.
\newblock \emph{\bibinfo{journal}{Nano Lett.}} \textbf{\bibinfo{volume}{22}},
  \bibinfo{pages}{8502--8508} (\bibinfo{year}{2022}).

\bibitem{NbVCoVTaFFSDE_NN2022}
\bibinfo{author}{{Narita}, H.} \emph{et~al.}
\newblock \bibinfo{title}{{Field-free superconducting diode effect in
  noncentrosymmetric superconductor/ferromagnet multilayers}}.
\newblock \emph{\bibinfo{journal}{Nat. Nanotechnol.}}
  \textbf{\bibinfo{volume}{17}}, \bibinfo{pages}{823--828}
  (\bibinfo{year}{2022}).

\bibitem{InAsSQUID_arXiv2023}
\bibinfo{author}{{Ciaccia}, C.} \emph{et~al.}
\newblock \bibinfo{title}{{Gate tunable {J}osephson diode in proximitized
  {InAs} supercurrent interferometers}}.
\newblock \emph{\bibinfo{journal}{Phys. Rev. Res.}}
  \textbf{\bibinfo{volume}{5}}, \bibinfo{pages}{033131} (\bibinfo{year}{2023}).

\bibitem{TiSQUIDSDE_PaolucciF_APL2023}
\bibinfo{author}{{Paolucci}, F.}, \bibinfo{author}{{De Simoni}, G.} \&
  \bibinfo{author}{{Giazotto}, F.}
\newblock \bibinfo{title}{{A gate- and flux-controlled supercurrent diode
  effect}}.
\newblock \emph{\bibinfo{journal}{Appl. Phys. Lett.}}
  \textbf{\bibinfo{volume}{122}}, \bibinfo{pages}{042601}
  (\bibinfo{year}{2023}).

\bibitem{JDEInGaAs3terminal_NC2023}
\bibinfo{author}{{Gupta}, M.} \emph{et~al.}
\newblock \bibinfo{title}{{Gate-tunable superconducting diode effect in a
  three-terminal {J}osephson device}}.
\newblock \emph{\bibinfo{journal}{Nat. Commun.}} \textbf{\bibinfo{volume}{14}},
  \bibinfo{pages}{3078} (\bibinfo{year}{2023}).

\bibitem{2DgasSQUIDdiode_arXiv2023}
\bibinfo{author}{{Valentini}, M.} \emph{et~al.}
\newblock \bibinfo{title}{{Parity-conserving Cooper-pair transport and ideal
  superconducting diode in planar germanium}}.
\newblock \emph{\bibinfo{journal}{Nat. Commun.}} \textbf{\bibinfo{volume}{15}},
  \bibinfo{pages}{169} (\bibinfo{year}{2024}).

\bibitem{Nb3Cl8FFSDE_WH_Nature2022}
\bibinfo{author}{{Wu}, H.} \emph{et~al.}
\newblock \bibinfo{title}{{The field-free {Josephson diode in a van der Waals}
  heterostructure}}.
\newblock \emph{\bibinfo{journal}{Nature}} \textbf{\bibinfo{volume}{604}},
  \bibinfo{pages}{653--656} (\bibinfo{year}{2022}).

\bibitem{FieldfreeSDE_NC2022}
\bibinfo{author}{{Golod}, T.} \& \bibinfo{author}{{Krasnov}, V.~M.}
\newblock \bibinfo{title}{{Demonstration of a superconducting
  diode-with-memory, operational at zero magnetic field with switchable
  nonreciprocity}}.
\newblock \emph{\bibinfo{journal}{Nat. Commun.}} \textbf{\bibinfo{volume}{13}},
  \bibinfo{pages}{3658} (\bibinfo{year}{2022}).

\bibitem{VFilm_HouYS_arXiv2022}
\bibinfo{author}{{Hou}, Y.} \emph{et~al.}
\newblock \bibinfo{title}{{Ubiquitous superconducting diode effect in
  superconductor thin films}}.
\newblock \emph{\bibinfo{journal}{Phys. Rev. Lett.}}
  \textbf{\bibinfo{volume}{131}}, \bibinfo{pages}{027001}
  (\bibinfo{year}{2023}).

\bibitem{JDEmageticAtton_TrahmsM_nature2023}
\bibinfo{author}{{Trahms}, M.} \emph{et~al.}
\newblock \bibinfo{title}{{Diode effect in Josephson junctions with a single
  magnetic atom}}.
\newblock \emph{\bibinfo{journal}{Nature}} \textbf{\bibinfo{volume}{615}},
  \bibinfo{pages}{628--633} (\bibinfo{year}{2023}).

\bibitem{DiamagneticSDE_SundareshA_NC2023}
\bibinfo{author}{{Sundaresh}, A.}, \bibinfo{author}{{V{\"a}yrynen}, J.~I.},
  \bibinfo{author}{{Lyanda-Geller}, Y.} \& \bibinfo{author}{{Rokhinson}, L.~P.}
\newblock \bibinfo{title}{{Diamagnetic mechanism of critical current
  non-reciprocity in multilayered superconductors}}.
\newblock \emph{\bibinfo{journal}{Nat. Commun.}} \textbf{\bibinfo{volume}{14}},
  \bibinfo{pages}{1628} (\bibinfo{year}{2023}).

\bibitem{Phi0sysmmetry_np2016}
\bibinfo{author}{{Szombati}, D.~B.} \emph{et~al.}
\newblock \bibinfo{title}{{Josephson {\ensuremath{\phi}}$_{0}$-Junction in
  nanowire quantum dots}}.
\newblock \emph{\bibinfo{journal}{Nat. Phys.}} \textbf{\bibinfo{volume}{12}},
  \bibinfo{pages}{568--572} (\bibinfo{year}{2016}).

\bibitem{AJEinNW_PRB2014}
\bibinfo{author}{{Yokoyama}, T.}, \bibinfo{author}{{Eto}, M.} \&
  \bibinfo{author}{{Nazarov}, Y.~V.}
\newblock \bibinfo{title}{{Anomalous Josephson effect induced by spin-orbit
  interaction and Zeeman effect in semiconductor nanowires}}.
\newblock \emph{\bibinfo{journal}{Phys. Rev. B}} \textbf{\bibinfo{volume}{89}},
  \bibinfo{pages}{195407} (\bibinfo{year}{2014}).

\bibitem{4piPeriodJJ_NC2016}
\bibinfo{author}{{Wiedenmann}, J.} \emph{et~al.}
\newblock \bibinfo{title}{{4{\ensuremath{\pi}}-periodic {Josephson supercurrent
  in HgTe-based topological Josephson} junctions}}.
\newblock \emph{\bibinfo{journal}{Nat. Commun.}} \textbf{\bibinfo{volume}{7}},
  \bibinfo{pages}{10303} (\bibinfo{year}{2016}).

\bibitem{EdgeSQUID_np2014}
\bibinfo{author}{{Hart}, S.} \emph{et~al.}
\newblock \bibinfo{title}{{Induced superconductivity in the quantum spin {H}all
  edge}}.
\newblock \emph{\bibinfo{journal}{Nat. Phys.}} \textbf{\bibinfo{volume}{10}},
  \bibinfo{pages}{638--643} (\bibinfo{year}{2014}).

\bibitem{EdgeSC_NN2015}
\bibinfo{author}{{Pribiag}, V.~S.} \emph{et~al.}
\newblock \bibinfo{title}{{Edge-mode superconductivity in a two-dimensional
  topological insulator}}.
\newblock \emph{\bibinfo{journal}{Nat. Nanotechnol.}}
  \textbf{\bibinfo{volume}{10}}, \bibinfo{pages}{593--597}
  (\bibinfo{year}{2015}).

\bibitem{JDESQUID_Fominov_PRB2022}
\bibinfo{author}{{Fominov}, Y.~V.} \& \bibinfo{author}{{Mikhailov}, D.~S.}
\newblock \bibinfo{title}{{Asymmetric higher-harmonic {SQUID} as a {J}osephson
  diode}}.
\newblock \emph{\bibinfo{journal}{Phys. Rev. B}}
  \textbf{\bibinfo{volume}{106}}, \bibinfo{pages}{134514}
  (\bibinfo{year}{2022}).

\bibitem{Ta2Pd3Te5QSH_GZP_PRB21}
\bibinfo{author}{{Guo}, Z.} \emph{et~al.}
\newblock \bibinfo{title}{{Quantum spin {Hall effect in Ta$_{2}$M$_{3}$Te$_{5}$
  (M =Pd ,Ni )}}}.
\newblock \emph{\bibinfo{journal}{Phys. Rev. B}}
  \textbf{\bibinfo{volume}{103}}, \bibinfo{pages}{115145}
  (\bibinfo{year}{2021}).

\bibitem{Ta2Pd3Te5TI_GuoZP_npjqm2022}
\bibinfo{author}{{Guo}, Z.}, \bibinfo{author}{{Deng}, J.},
  \bibinfo{author}{{Xie}, Y.} \& \bibinfo{author}{{Wang}, Z.}
\newblock \bibinfo{title}{{Quadrupole topological insulators in
  {Ta$_{2}$M$_{3}$Te$_{5}$ (M = Ni, Pd)} monolayers}}.
\newblock \emph{\bibinfo{journal}{npj Quant. Mater.}}
  \textbf{\bibinfo{volume}{7}}, \bibinfo{pages}{87} (\bibinfo{year}{2022}).

\bibitem{Ta2Pd3Te5LL_arXiv2022}
\bibinfo{author}{{Wang}, A.} \emph{et~al.}
\newblock \bibinfo{title}{{A robust and tunable {Luttinger liquid in correlated
  edge of transition-metal second-order topological insulator
  Ta$_2$Pd$_3$Te$_5$}}}.
\newblock \emph{\bibinfo{journal}{Nat. Commun.}} \textbf{\bibinfo{volume}{14}},
  \bibinfo{pages}{7647} (\bibinfo{year}{2023}).

\bibitem{Ta2Pd3Te5ExcitionIn_PRX2024}
\bibinfo{author}{Huang, J.} \emph{et~al.}
\newblock \bibinfo{title}{Evidence for an excitonic insulator state in
  {Ta$_{2}$Pd$_{3}$Te$_{5}$}}.
\newblock \emph{\bibinfo{journal}{Phys. Rev. X}} \textbf{\bibinfo{volume}{14}},
  \bibinfo{pages}{011046} (\bibinfo{year}{2024}).

\bibitem{Ta2Pd3Te5ExcitionEdge_arXiv2023}
\bibinfo{author}{{Shafayat Hossain}, M.} \emph{et~al.}
\newblock \bibinfo{title}{{Discovery of a topological exciton insulator with
  tunable momentum order}}.
\newblock \emph{\bibinfo{journal}{arXiv:2312.15862}}.

\bibitem{Ta2Pd3Te5ExcitionIn2_PRX2024}
\bibinfo{author}{Zhang, P.} \emph{et~al.}
\newblock \bibinfo{title}{Spontaneous gap opening and potential excitonic
  states in an ideal {D}irac semimetal {Ta$_{2}$Pd$_{3}$Te$_{5}$}}.
\newblock \emph{\bibinfo{journal}{Phys. Rev. X}} \textbf{\bibinfo{volume}{14}},
  \bibinfo{pages}{011047} (\bibinfo{year}{2024}).

\bibitem{Ta2Pd3Te5STM_WangXG_PRB2021}
\bibinfo{author}{{Wang}, X.} \emph{et~al.}
\newblock \bibinfo{title}{{Observation of topological edge states in the
  quantum spin {Hall insulator Ta$_{2}$Pd$_{3}$Te$_{5}$}}}.
\newblock \emph{\bibinfo{journal}{Phys. Rev. B}}
  \textbf{\bibinfo{volume}{104}}, \bibinfo{pages}{L241408}
  (\bibinfo{year}{2021}).

\bibitem{Ta2Pd3Te5thermometer_Arxiv24}
\bibinfo{author}{{Li}, Y.} \emph{et~al.}
\newblock \bibinfo{title}{{Ta$_{2}$Pd$_{3}$Te$_{5}$ topological thermometer}}.
\newblock \emph{\bibinfo{journal}{arXiv: 2406.00959}}.

\bibitem{Ta2Pd3Te5dopingSC_JPSJ2021}
\bibinfo{author}{{Higashihara}, N.} \emph{et~al.}
\newblock \bibinfo{title}{{Superconductivity in {Nb$_{2}$Pd$_{3}$Te$_{5}$ and
  chemically-doped Ta$_{2}$Pd$_{3}$Te$_{5}$}}}.
\newblock \emph{\bibinfo{journal}{J. Phys. Soc. Jpn.}}
  \textbf{\bibinfo{volume}{90}}, \bibinfo{pages}{063705}
  (\bibinfo{year}{2021}).

\bibitem{Ta2Pd3Te5presureSC_arXiv2023}
\bibinfo{author}{{Yu}, H.} \emph{et~al.}
\newblock \bibinfo{title}{{Observation of emergent superconductivity in the
  quantum spin Hall insulator Ta$_{2}$Pd$_{3}$Te$_{5}$ via pressure
  manipulation}}.
\newblock \emph{\bibinfo{journal}{J. Am. Chem. Soc.}}
  \textbf{\bibinfo{volume}{146}}, \bibinfo{pages}{3890--3899}
  (\bibinfo{year}{2024}).

\bibitem{Ta2Ni3Te5pressureSC_YangHY_PRB2023}
\bibinfo{author}{{Yang}, H.} \emph{et~al.}
\newblock \bibinfo{title}{{Pressure-induced nontrivial {Z$_{2}$ band topology
  and superconductivity in the transition metal chalcogenide
  Ta$_{2}$Ni$_{3}$Te$_{5}$}}}.
\newblock \emph{\bibinfo{journal}{Phys. Rev. B}}
  \textbf{\bibinfo{volume}{107}}, \bibinfo{pages}{L020503}
  (\bibinfo{year}{2023}).

\bibitem{InSbSnJDE_arXiv2022}
\bibinfo{author}{{Zhang}, B.} \emph{et~al.}
\newblock \bibinfo{title}{{Evidence of $\phi$0-Josephson junction from skewed
  diffraction patterns in Sn-InSb nanowires}}.
\newblock \emph{\bibinfo{journal}{arXiv:2212.00199}}.

\bibitem{InSbnanowire_arXiv2022}
\bibinfo{author}{{Mazur}, G.~P.} \emph{et~al.}
\newblock \bibinfo{title}{{The gate-tunable Josephson diode}}.
\newblock \emph{\bibinfo{journal}{arXiv:2211.14283}}.

\bibitem{TriodeGraphene_NanoL2023}
\bibinfo{author}{{Chiles}, J.} \emph{et~al.}
\newblock \bibinfo{title}{{Nonreciprocal supercurrents in a field-free graphene
  Josephson triode}}.
\newblock \emph{\bibinfo{journal}{Nano Lett.}} \textbf{\bibinfo{volume}{23}},
  \bibinfo{pages}{5257--5263} (\bibinfo{year}{2023}).

\bibitem{BismuthHOTIsquid_NC2017}
\bibinfo{author}{{Murani}, A.} \emph{et~al.}
\newblock \bibinfo{title}{{Ballistic edge states in {Bismuth nanowires revealed
  by SQUID} interferometry}}.
\newblock \emph{\bibinfo{journal}{Nat. Commun.}} \textbf{\bibinfo{volume}{8}},
  \bibinfo{pages}{15941} (\bibinfo{year}{2017}).

\bibitem{CrackinmonolayerWTe2_SA2019}
\bibinfo{author}{{Shi}, Y.} \emph{et~al.}
\newblock \bibinfo{title}{{Imaging quantum spin Hall edges in monolayer
  WTe$_{2}$}}.
\newblock \emph{\bibinfo{journal}{Sci. Adv.}} \textbf{\bibinfo{volume}{5}},
  \bibinfo{pages}{eaat8799} (\bibinfo{year}{2019}).

\bibitem{2DSC_SatioY_NRM2017}
\bibinfo{author}{{Saito}, Y.}, \bibinfo{author}{{Nojima}, T.} \&
  \bibinfo{author}{{Iwasa}, Y.}
\newblock \bibinfo{title}{{Highly crystalline 2{D} superconductors}}.
\newblock \emph{\bibinfo{journal}{Nat. Rev. Mater.}}
  \textbf{\bibinfo{volume}{2}}, \bibinfo{pages}{16094} (\bibinfo{year}{2017}).

\bibitem{FST_Chauvin_PRL2006}
\bibinfo{author}{{Chauvin}, M.} \emph{et~al.}
\newblock \bibinfo{title}{{Superconducting atomic contacts under microwave
  irradiation}}.
\newblock \emph{\bibinfo{journal}{Phys. Rev. Lett.}}
  \textbf{\bibinfo{volume}{97}}, \bibinfo{pages}{067006}
  (\bibinfo{year}{2006}).

\bibitem{InAsAlSQUIDjde_arXiv2023}
\bibinfo{author}{{Ciaccia}, C.} \emph{et~al.}
\newblock \bibinfo{title}{{Charge-4e supercurrent in a two-dimensional
  {InAs-Al} superconductor-semiconductor heterostructure}}.
\newblock \emph{\bibinfo{journal}{Commun. Phys.}} \textbf{\bibinfo{volume}{7}},
  \bibinfo{pages}{41} (\bibinfo{year}{2024}).

\bibitem{Al2DEG_BaumgartnerC_NN2022}
\bibinfo{author}{{Baumgartner}, C.} \emph{et~al.}
\newblock \bibinfo{title}{{Supercurrent rectification and magnetochiral effects
  in symmetric {J}osephson junctions}}.
\newblock \emph{\bibinfo{journal}{Nat. Nanotechnol.}}
  \textbf{\bibinfo{volume}{17}}, \bibinfo{pages}{39--44}
  (\bibinfo{year}{2022}).

\bibitem{Qubit_PRL2020}
\bibinfo{author}{{Larsen}, T.~W.} \emph{et~al.}
\newblock \bibinfo{title}{{Parity-protected superconductor-semiconductor
  qubit}}.
\newblock \emph{\bibinfo{journal}{Phys. Rev. Lett.}}
  \textbf{\bibinfo{volume}{125}}, \bibinfo{pages}{056801}
  (\bibinfo{year}{2020}).

\bibitem{SCQubit_PRXQ2022}
\bibinfo{author}{{Schrade}, C.}, \bibinfo{author}{{Marcus}, C.~M.} \&
  \bibinfo{author}{{Gyenis}, A.}
\newblock \bibinfo{title}{{Protected hybrid superconducting qubit in an array
  of gate-tunable Josephson interferometers}}.
\newblock \emph{\bibinfo{journal}{PRX Quantum}} \textbf{\bibinfo{volume}{3}},
  \bibinfo{pages}{030303} (\bibinfo{year}{2022}).

\bibitem{SCQubit2_PRXQ2022}
\bibinfo{author}{{Maiani}, A.}, \bibinfo{author}{{Kjaergaard}, M.} \&
  \bibinfo{author}{{Schrade}, C.}
\newblock \bibinfo{title}{{Entangling transmons with low-frequency protected
  superconducting qubits}}.
\newblock \emph{\bibinfo{journal}{PRX Quantum}} \textbf{\bibinfo{volume}{3}},
  \bibinfo{pages}{030329} (\bibinfo{year}{2022}).

\bibitem{FractionalShapiro_PRB2020}
\bibinfo{author}{{Raes}, B.} \emph{et~al.}
\newblock \bibinfo{title}{{Fractional shapiro steps in resistively shunted
  Josephson junctions as a fingerprint of a skewed current-phase
  relationship}}.
\newblock \emph{\bibinfo{journal}{Phys. Rev. B}}
  \textbf{\bibinfo{volume}{102}}, \bibinfo{pages}{054507}
  (\bibinfo{year}{2020}).

\bibitem{MoS2NCT_WakatsukiR_SA2017}
\bibinfo{author}{{Wakatsuki}, R.} \emph{et~al.}
\newblock \bibinfo{title}{{Nonreciprocal charge transport in noncentrosymmetric
  superconductors}}.
\newblock \emph{\bibinfo{journal}{Sci. Adv.}} \textbf{\bibinfo{volume}{3}},
  \bibinfo{pages}{e1602390} (\bibinfo{year}{2017}).

\bibitem{NTin2DSC_PRB2018}
\bibinfo{author}{{Hoshino}, S.}, \bibinfo{author}{{Wakatsuki}, R.},
  \bibinfo{author}{{Hamamoto}, K.} \& \bibinfo{author}{{Nagaosa}, N.}
\newblock \bibinfo{title}{{Nonreciprocal charge transport in two-dimensional
  noncentrosymmetric superconductors}}.
\newblock \emph{\bibinfo{journal}{Phys. Rev. B}} \textbf{\bibinfo{volume}{98}},
  \bibinfo{pages}{054510} (\bibinfo{year}{2018}).

\bibitem{NanotubeLPR2w_NC2017}
\bibinfo{author}{{Qin}, F.} \emph{et~al.}
\newblock \bibinfo{title}{{Superconductivity in a chiral nanotube}}.
\newblock \emph{\bibinfo{journal}{Nat. Commun.}} \textbf{\bibinfo{volume}{8}},
  \bibinfo{pages}{14465} (\bibinfo{year}{2017}).

\bibitem{SrTiO3GateNCT_ItahashiYM_SA2020}
\bibinfo{author}{{Itahashi}, Y.~M.} \emph{et~al.}
\newblock \bibinfo{title}{{Nonreciprocal transport in gate-induced polar
  superconductor {SrTiO}$_{3}$}}.
\newblock \emph{\bibinfo{journal}{Sci. Adv.}} \textbf{\bibinfo{volume}{6}},
  \bibinfo{pages}{eaay9120} (\bibinfo{year}{2020}).

\bibitem{SrTiO3LaTiO3nonreciprocal_NC2019}
\bibinfo{author}{{Choe}, D.} \emph{et~al.}
\newblock \bibinfo{title}{{Gate-tunable giant nonreciprocal charge transport in
  noncentrosymmetric oxide interfaces}}.
\newblock \emph{\bibinfo{journal}{Nat. Commun.}} \textbf{\bibinfo{volume}{10}},
  \bibinfo{pages}{4510} (\bibinfo{year}{2019}).

\bibitem{BiSbTe3_LeggHF_NN2022}
\bibinfo{author}{{Legg}, H.~F.} \emph{et~al.}
\newblock \bibinfo{title}{{Giant magnetochiral anisotropy from quantum-confined
  surface states of topological insulator nanowires}}.
\newblock \emph{\bibinfo{journal}{Nat. Nanotechnol.}}
  \textbf{\bibinfo{volume}{17}}, \bibinfo{pages}{696--700}
  (\bibinfo{year}{2022}).

\bibitem{NbSe2NCT_ZhangEZ_NC2020}
\bibinfo{author}{{Zhang}, E.} \emph{et~al.}
\newblock \bibinfo{title}{{Nonreciprocal superconducting {NbSe}$_{2}$
  antenna}}.
\newblock \emph{\bibinfo{journal}{Nat. Commun.}} \textbf{\bibinfo{volume}{11}},
  \bibinfo{pages}{5634} (\bibinfo{year}{2020}).

\bibitem{Dilutionrefrigerator_Cryogenics2022}
\bibinfo{author}{{Zu}, H.}, \bibinfo{author}{{Dai}, W.} \& \bibinfo{author}{{de
  Waele}, A.~T.~A.~M.}
\newblock \bibinfo{title}{{Development of dilution refrigerators-{a} review}}.
\newblock \emph{\bibinfo{journal}{Cryogenics}} \textbf{\bibinfo{volume}{121}},
  \bibinfo{pages}{103390} (\bibinfo{year}{2022}).

\bibitem{RuO2downto5mK_Cryogenics2021}
\bibinfo{author}{{Myers}, S.~A.}, \bibinfo{author}{{Li}, H.} \&
  \bibinfo{author}{{Cs{\'a}thy}, G.~A.}
\newblock \bibinfo{title}{{A ruthenium oxide thermometer for dilution
  refrigerators operating down to 5 mK}}.
\newblock \emph{\bibinfo{journal}{Cryogenics}} \textbf{\bibinfo{volume}{119}},
  \bibinfo{pages}{103367} (\bibinfo{year}{2021}).

\bibitem{ThermoR_AIPCP2002}
\bibinfo{author}{{Yeager}, C.~J.}, \bibinfo{author}{{Courts}, S.~S.} \&
  \bibinfo{author}{{Davenport}, W.~E.}
\newblock \bibinfo{title}{{Thermal resistance of cryogenic thermometers at
  ultra-low temperatures}}.
\newblock \emph{\bibinfo{journal}{AIP Conf. Proc.}}
  \textbf{\bibinfo{volume}{47}}, \bibinfo{pages}{1644--1650}
  (\bibinfo{year}{2002}).

\bibitem{NernstinWTe2_arXiv23}
\bibinfo{author}{{Song}, T.} \emph{et~al.}
\newblock \bibinfo{title}{{Unconventional superconducting quantum criticality
  in monolayer {WTe$_{2}$}}}.
\newblock \emph{\bibinfo{journal}{Nat. Phys.}} \textbf{\bibinfo{volume}{20}},
  \bibinfo{pages}{269--274} (\bibinfo{year}{2024}).

\end{thebibliography}
\end{document}


\title{Supplementary Information for \\
Interfering Josephson diode effect in Ta$_{2}$Pd$_{3}$Te$_{5}$ asymmetric edge interferometer }

\author{Yupeng Li}
      \thanks{Equal contributions}
      \affiliation{Beijing National Laboratory for Condensed Matter Physics, Institute of Physics, Chinese Academy of Sciences, Beijing 100190, China}

\author{Dayu Yan}
      \thanks{Equal contributions}
      \affiliation{Beijing National Laboratory for Condensed Matter Physics, Institute of Physics, Chinese Academy of Sciences, Beijing 100190, China}

\author{Yu Hong}
      \affiliation{Beijing National Laboratory for Condensed Matter Physics, Institute of Physics, Chinese Academy of Sciences, Beijing 100190, China}
      \affiliation{School of Physical Sciences, University of Chinese Academy of Sciences, Beijing 100049, China}

\author{Haohao Sheng}
      \affiliation{Beijing National Laboratory for Condensed Matter Physics, Institute of Physics, Chinese Academy of Sciences, Beijing 100190, China}
      \affiliation{School of Physical Sciences, University of Chinese Academy of Sciences, Beijing 100049, China}

\author{Anqi Wang}
      \affiliation{Beijing National Laboratory for Condensed Matter Physics, Institute of Physics, Chinese Academy of Sciences, Beijing 100190, China}
      \affiliation{School of Physical Sciences, University of Chinese Academy of Sciences, Beijing 100049, China}

\author{Ziwei Dou}
      \affiliation{Beijing National Laboratory for Condensed Matter Physics, Institute of Physics, Chinese Academy of Sciences, Beijing 100190, China}

\author{Xingchen Guo}
      \affiliation{Beijing National Laboratory for Condensed Matter Physics, Institute of Physics, Chinese Academy of Sciences, Beijing 100190, China}
      \affiliation{School of Physical Sciences, University of Chinese Academy of Sciences, Beijing 100049, China}

\author{Xiaofan Shi}
      \affiliation{Beijing National Laboratory for Condensed Matter Physics, Institute of Physics, Chinese Academy of Sciences, Beijing 100190, China}
      \affiliation{School of Physical Sciences, University of Chinese Academy of Sciences, Beijing 100049, China}

\author{Zikang Su}
      \affiliation{Beijing National Laboratory for Condensed Matter Physics, Institute of Physics, Chinese Academy of Sciences, Beijing 100190, China}
      \affiliation{School of Physical Sciences, University of Chinese Academy of Sciences, Beijing 100049, China}

\author{Zhaozheng Lyu}
      \affiliation{Beijing National Laboratory for Condensed Matter Physics, Institute of Physics, Chinese Academy of Sciences, Beijing 100190, China}

\author{Tian Qian}
      \affiliation{Beijing National Laboratory for Condensed Matter Physics, Institute of Physics, Chinese Academy of Sciences, Beijing 100190, China}
      \affiliation{Songshan Lake Materials Laboratory, Dongguan 523808, China}

\author{Guangtong Liu}
      \affiliation{Beijing National Laboratory for Condensed Matter Physics, Institute of Physics, Chinese Academy of Sciences, Beijing 100190, China}
      \affiliation{Songshan Lake Materials Laboratory, Dongguan 523808, China}

\author{Fanming Qu}
      \affiliation{Beijing National Laboratory for Condensed Matter Physics, Institute of Physics, Chinese Academy of Sciences, Beijing 100190, China}
      \affiliation{School of Physical Sciences, University of Chinese Academy of Sciences, Beijing 100049, China}
      \affiliation{Songshan Lake Materials Laboratory, Dongguan 523808, China}

\author{Kun Jiang}
      \affiliation{Beijing National Laboratory for Condensed Matter Physics, Institute of Physics, Chinese Academy of Sciences, Beijing 100190, China}

\author{Zhijun Wang}
      \affiliation{Beijing National Laboratory for Condensed Matter Physics, Institute of Physics, Chinese Academy of Sciences, Beijing 100190, China}
      \affiliation{School of Physical Sciences, University of Chinese Academy of Sciences, Beijing 100049, China}

\author{Youguo Shi}
      \email{ygshi@iphy.ac.cn}
      \affiliation{Beijing National Laboratory for Condensed Matter Physics, Institute of Physics, Chinese Academy of Sciences, Beijing 100190, China}
      \affiliation{Songshan Lake Materials Laboratory, Dongguan 523808, China}

\author{Zhu-An Xu}
      \affiliation{School of Physics, Zhejiang University, Hangzhou, China}
      \affiliation{State Key Laboratory of Silicon and Advanced Semiconductor Materials, Zhejiang University, Hangzhou, China}
      \affiliation{Hefei National Laboratory, Hefei 230088, China}

\author{Jiangping Hu}
      \affiliation{Beijing National Laboratory for Condensed Matter Physics, Institute of Physics, Chinese Academy of Sciences, Beijing 100190, China}
      \affiliation{Kavli Institute of Theoretical Sciences, University of Chinese Academy of Sciences, Beijing 100190, China}

\author{Li Lu}
      \email{lilu@iphy.ac.cn}
      \affiliation{Beijing National Laboratory for Condensed Matter Physics, Institute of Physics, Chinese Academy of Sciences, Beijing 100190, China}
      \affiliation{School of Physical Sciences, University of Chinese Academy of Sciences, Beijing 100049, China}
      \affiliation{Songshan Lake Materials Laboratory, Dongguan 523808, China}

\author{Jie Shen}
      \email{shenjie@iphy.ac.cn}
      \affiliation{Beijing National Laboratory for Condensed Matter Physics, Institute of Physics, Chinese Academy of Sciences, Beijing 100190, China}
      \affiliation{Songshan Lake Materials Laboratory, Dongguan 523808, China}
      \affiliation{Beijing Academy of Quantum Information Sciences, Beijing 100193, China}


\maketitle
\noindent\textbf{Contents:}


\noindent\textbf{Supplementary Note 1. Superconductivity of both Ta$_{2}$Pd$_{3}$Te$_{5}$ devices}

\noindent\textbf{Supplementary Note 2. Three requirements to form JDE in the interferometer}

\noindent\textbf{Supplementary Note 3. Estimation of switching current for two devices}

\noindent\textbf{Supplementary Note 4. Temperature-dependent JDE}

\noindent\textbf{Supplementary Note 5. Excluding effects arising from finite-momentum superconductor}

\noindent\textbf{Supplementary Note 6. Fractional Shapiro steps under microwave}

\noindent\textbf{Supplementary Note 7. Antisymmetric second harmonic transport of both devices}

\noindent\textbf{Supplementary Note 8. Discussion on self-inductance of interferometer}

\noindent\textbf{Supplementary Note 9. Analysis of supercurrent density profile}

\vspace{3ex}

\noindent\textbf{Supplementary Note 1. Superconductivity of both Ta$_{2}$Pd$_{3}$Te$_{5}$ devices}

In Supplementary Figure 1, the induced superconducting transition temperature ($T_{c}$) is
0.46 K and 0.51 K for devices S1 and S2, respectively,
determined by a 50\% drop in normal resistivity.
The zero-temperature induced superconducting gap is $\Delta$ = 1.76 $k_{B}T_{c}$ = 0.07 meV
(0.08 meV) for device S1 (device S2).

The transparency of an interface or junction is commonly measured to describe the contact quality
and barrier strength ($Z$) \cite{BKTtranmision_PRB1982,4piPeriodJJ_NC2016,fractionalAC_NP2012}.
The excess current $I_{exc}$ is commonly applied to estimate the transparency $\tau_{SN}$
of an SN interface according to BKT theory \cite{BKTtranmision_PRB1982}.
In this case, $I_{exc}$ is estimated to be 0.076 $\mu$A (0.336 $\mu$A) from $I_{b}$ - $V$ curve
in Supplementary Figure 1b for device S1 (device S2).
For device S1 (device S2), we have $eI_{exc}R_{N}$/$\Delta$ = 0.344 (0.242) \cite{SISINISIS_Transparency_APL2019,SISINISIS_Transparency1_PRA2017,fractionalAC_NP2012},
which roughly corresponds to $\tau$ = $\tau_{SN}^{2}$ = $1/(1 + Z^{2})^2$ = 0.230 (0.182) for the SNS junction \cite{OBTK_PRB1983,FractionalACJ_arXiv2021}.
This transparency is in agreement with the observed experiment phenomena
(sawtooth-shaped $I_{c}$ pattern and fractional Shapiro steps).
It should be noted that the estimated $\tau$ represents the approximate total transparency in our JJs.
Superconducting rectification effect in device S1 is shown in Supplementary Figure 1c.

\begin{figure*}[!thb]
\begin{center}
\includegraphics[width=6.5in]{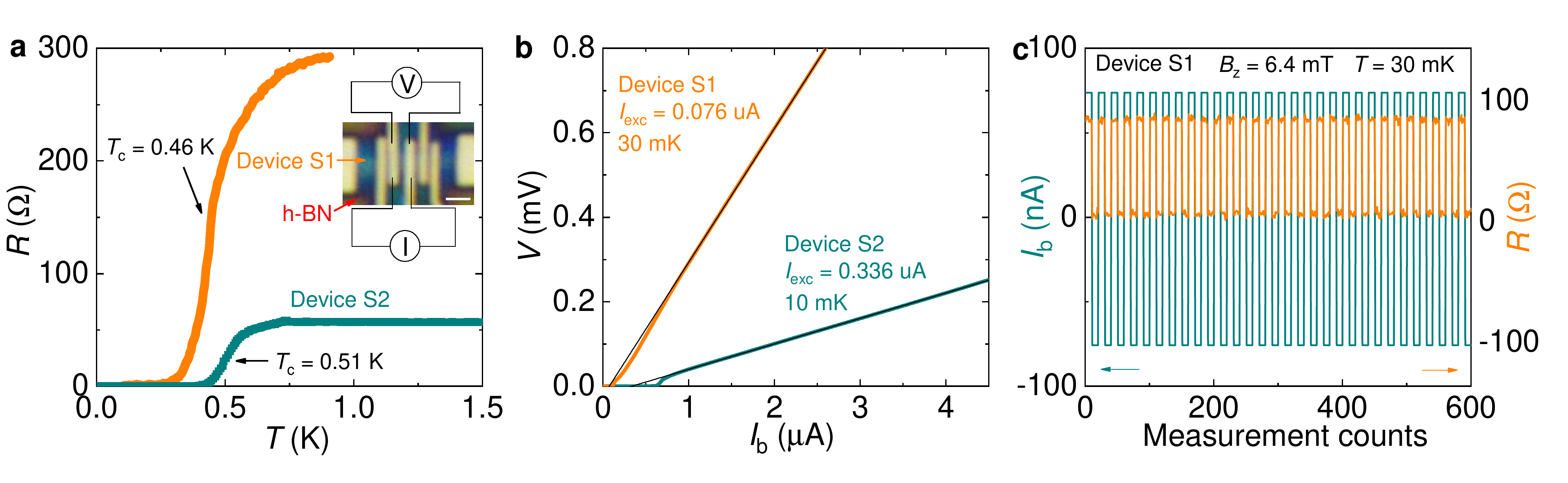}
\end{center}
\caption{\label{Fig4.5} \textbf{Superconductivity in both Ta$_{2}$Pd$_{3}$Te$_{5}$ devices.}
$\mathbf{a}$, Superconductivity of devices S1 and S2.
The inset is the photomicrograph of device S1 and the measurement configuration.
The white scale bar corresponds to 1 $\mu$m.
$\mathbf{b}$, $I$-$V$ curve for both devices.
The black asymptotes do not cross the origin, suggesting the presence of excess current for both devices.
$\mathbf{c}$, Superconducting rectification effect observed in device S1 with switching power $\sim$ 0.56 pW.
   }
\end{figure*}

\begin{figure*}[!thb]
\begin{center}
\includegraphics[width=6.5in]{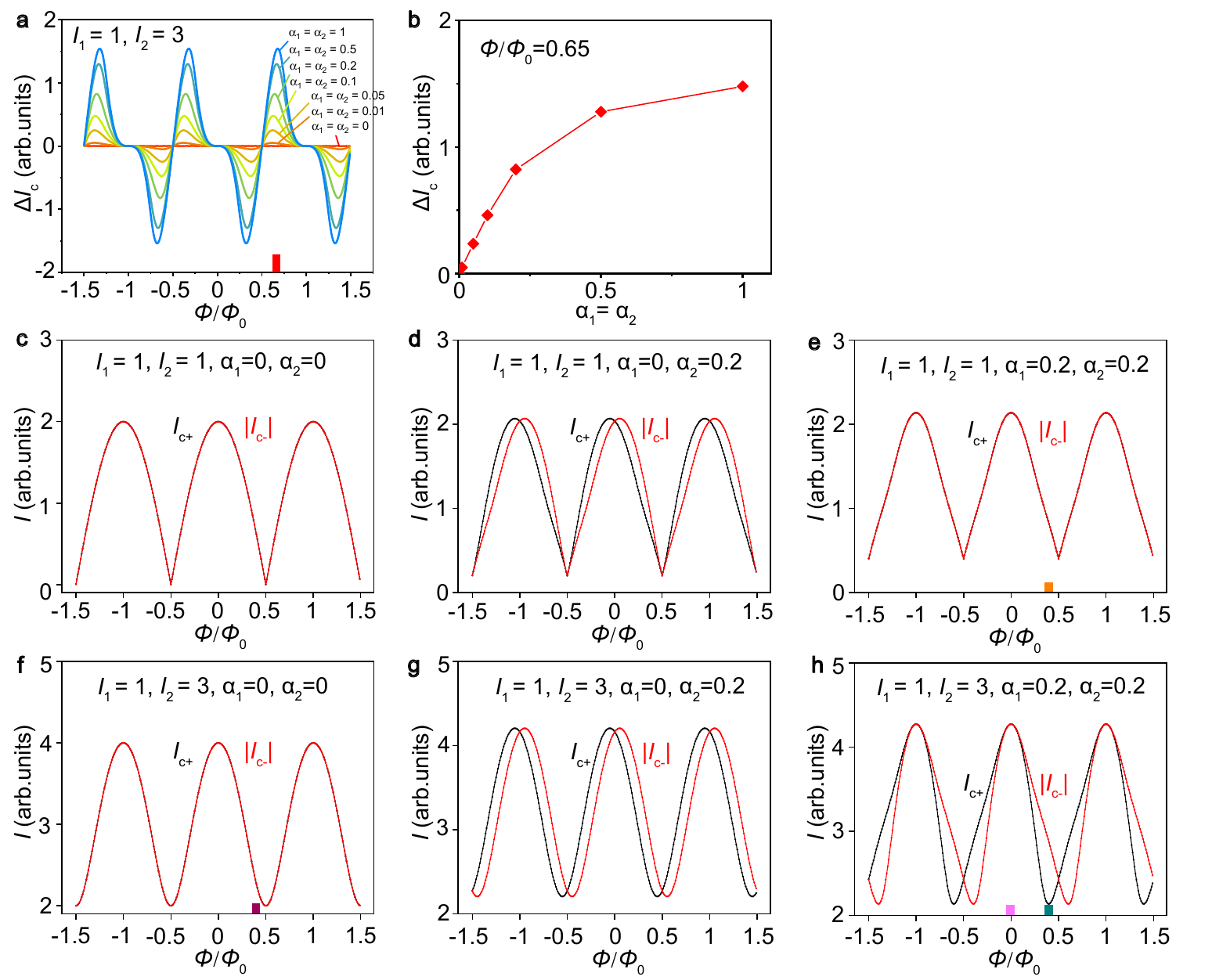}
\end{center}
\caption{\label{Fig6} \textbf{Simulated JDE for supercurrent interferometer with two JJs in parallel to indicate the
importance of unequal supercurrent, magnetic field and higher harmonic in the CPR.}
$\mathbf{a}$, $\Delta I_{c} = I_{c+} - \left|I_{c-}\right|$ as a function of flux at different $\alpha_{1}$ ($\alpha_{2}$).
$\mathbf{b}$, $\Delta I_{c}$ as a function of $\alpha_{1}$ and $\alpha_{2}$.
$\mathbf{c}$, No JDE for $I_{1}$ = 1, $I_{2}$ = 1 and $\alpha_{1}$ = $\alpha_{2}$ = 0.
$\mathbf{d}$, Flux-quantum-dependent JDE for $I_{1}$ = $I_{2}$ = 1, $\alpha_{1}$ = 0 and $\alpha_{2}$ = 0.2.
$\mathbf{e}$, No JDE for $I_{1}$ = $I_{2}$ = 1 and $\alpha_{1}$ = $\alpha_{2}$ = 0.2.
$\mathbf{f}$, No JDE for $I_{1}$ = 1, $I_{2}$ = 3 and $\alpha_{1}$ = $\alpha_{2}$ = 0.
$\mathbf{g}$, JDE for $I_{1}$ = 1, $I_{2}$ = 3, $\alpha_{1}$ = 0 and $\alpha_{2}$ = 0.2.
$\mathbf{h}$, JDE for $I_{1}$ = 1, $I_{2}$ = 3 and $\alpha_{1}$ = $\alpha_{2}$ = 0.2.
  }
\end{figure*}

\vspace{3ex}

\noindent\textbf{Supplementary Note 2. Three requirements to form JDE in the interferometer}

A minimal model with $n$ JJs in parallel can be used to approximatively describe
the Josephson diode effect (JDE) in the interferometer \cite{FDEinSI_Constanti_PRL2022}.
In this model, only the first and second harmonic contribution in the CPR are considered.
The supercurrent is written as
\begin{equation}\label{LK}
I (\varphi,\phi/\phi_{0}) = \sum_{1}^n I_{n}sin(\varphi + 2\pi\zeta_{n}\phi/\phi_{0}) + \alpha_{n}I_{n}sin(2\varphi + 4\pi\zeta_{n}\phi/\phi_{0}),
\end{equation}
where $\varphi$ is the superconducting phase difference,
$I_{n}$ ($\alpha_{n}I_{n}$) is the amplitude of the first (second) harmonic current for the $n$th JJ,
$\alpha_{n}$ is the ratio of the second harmonic to the first harmonic,
$\phi$ is the magnetic flux and $\phi_{0} = h/2e$ is the magnetic flux quantum.
$\zeta_{n}$ is the modified ratio,
and $\zeta_{1}$ = 0, $\zeta_{2}$ = 1 ($\zeta_{1}$ = 0, $\zeta_{2}$ = 1, $\zeta_{3}$ = 0.23)
for the simulation of the two (three) JJs case, where
$\zeta_{3}$ represents the enclosed-area ratio of two corresponding SQUID patterns in device S2.

For simplicity, the behavior of the interferometers and the emergence of JDE are explained using a SQUID formed by two JJs and three key requirements \cite{FDEinSI_Constanti_PRL2022}, such as unequal supercurrent, magnetic field and the high-order harmonic in the CPR.
First, JDE exists at $\alpha_{n} \neq 0$ and increases with increasing $\alpha_{n}$ in Supplementary Figure 2a, b.
Within this context, $\alpha_{n} = 0.2$ is chosen as an illustrative example for detailed introduction.
The CPR for different cases is illustrated in Figure 1c-f of the main text,
with critical current marked under corresponding conditions using the same color indicator as in Supplementary Figure 2.
A comparison with the cases in Supplementary Figure 2f, h suggests that the higher harmonic ($\alpha_{n} \neq 0$)
and magnetic field condictions ($\phi/\phi_{0} \neq 0$) are vital factors in inducing the JDE.
Additionally, in Supplementary Figure 2c, d, the unequal supercurrent (comprising the first and second harmonic contributions) is also an essential requirement for the emergence of JDE.

\begin{figure*}[!thb]
\begin{center}
\includegraphics[width=5.5in]{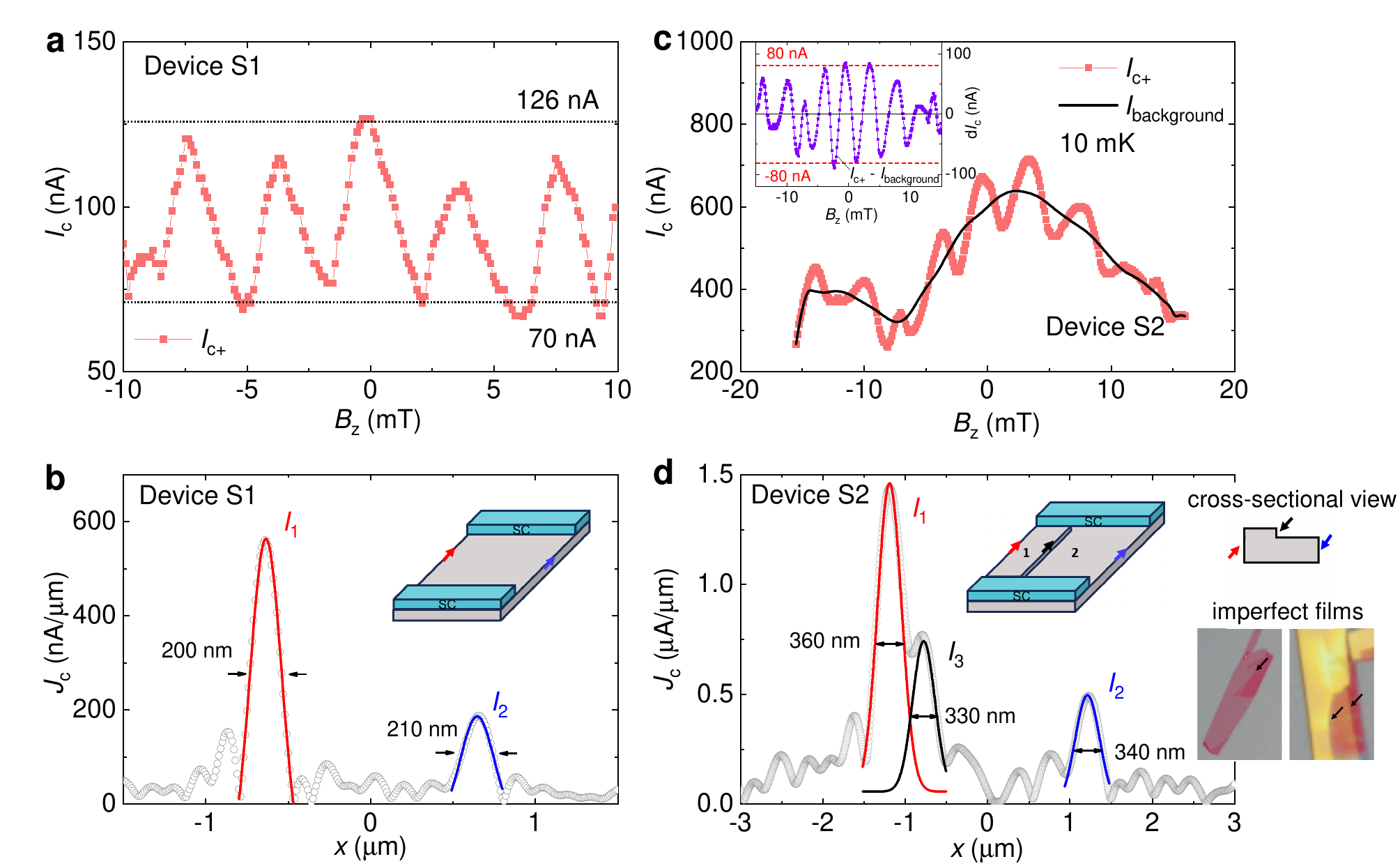}
\end{center}
\caption{\label{Fig8} \textbf{Estimation of switching current for different channels
in Ta$_{2}$Pd$_{3}$Te$_{5}$ devices S1 and S2.}
$\mathbf{a}$, Switching current $I_{c+}$ of device S1.
$\mathbf{b}$, Supercurrent density distribution of device S1.
The supercurrents originate from edges of the JJ,
and they can be well fitted by the Gauss function (red and blue lines).
$\mathbf{c}$, $I_{c+}$ for device S2.
$\mathbf{d}$, Supercurrent density distribution of device S2.
The supercurrent seems to originate from three edges (red, black and blue lines),
which can be viewed as three JJs in parallel.
The upper right insert is a cross-sectional view of device S2,
while the bottom insert shows examples of imperfect films with the middle edge in the bulk.
  }
\end{figure*}

\vspace{3ex}

\vspace{3ex}

\vspace{3ex}

\noindent\textbf{Supplementary Note 3. Estimation of switching current for two devices}

Before simulating JDE for our cases, we need to first estimate the switching currents of different channels in both Ta$_{2}$Pd$_{3}$Te$_{5}$ devices, in order to obtain initial values for simulation.
For device S1 in Supplementary Figure 3a, the oscillation of the switching current varies approximately between 126 nA and 70 nA
(short dotted lines) with a period of $\sim$ 3.7 mT (corresponding enclosed area of the JJ is calculated to be 0.56 $\mu m^{-2}$, close to the actual area 0.52 $\mu m^{-2}$).
Supercurrent density distribution in Supplementary Figure 3b shows that two edge supercurrents contribute to the SQUID pattern.
Because of the small number of periods about the SQUID pattern,
the signal of distribution in the bulk state seems similar to that of the outer state,
suggesting that they are not real states.
Correspondingly, two suppercurrent $I_{1}$ = 98 nA and $I_{2}$ = 28 nA are estimated from $I_{1} + I_{2}$ = 126 nA
and $I_{1} - I_{2}$ = 70 nA formulas, without taking the higher harmonic into consideration.
For device S2 in Supplementary Figure 3c, the SQUID pattern is quite asymmetric
and seems to show two sets of interference patterns,
which are caused by three interfering suppercurrents from edge states in the whole JJ (Supplementary Figure 3d).
The middle edge state maybe stem from the imperfect exfoliation,
forming a ladder-like structure comprised of the top layer 1 and bottom layer 2,
as shown in the inset of Supplementary Figure 3d.
Therefore, the Ta$_{2}$Pd$_{3}$Te$_{5}$ JJ of device S2 can be viewed as three JJs in parallel.
In Supplementary Figure 3c, the background (black line) of the total SQUID pattern approximately displays one period of
$\sim$ 18.9 mT with an enclosed area 0.109 $\mu m^{-2}$ formed by JJ$_{1}$ (or $I_{1}$) and JJ$_{3}$ (or $I_{3}$).
$I_{1} + I_{3}$ $\sim$ 636 nA and $I_{1} - I_{3}$ $\sim$ 320 nA are obtained.
The inset in Supplementary Figure 3c shows the main oscillation of $I_{2}$ after deducting the background,
and $I_{2}$ = 80 nA is approximately obtained with a period of $\sim$ 4.5 mT
(enclosed area 0.458 $\mu m^{-2}$ formed by JJ$_{1}$ and JJ$_{2}$, which is close to the actual area 0.466 $\mu m^{-2}$). Resultantly, $I_{1}$ = 478 nA and $I_{3}$ = 158 nA.
It is important to note that these estimations of $I_{n}$ are rough and serve as initial values for fitting,
because the higher harmonic actually exists in our devices, confirmed by microwave measurements.

\begin{figure*}[!thb]
\begin{center}
\includegraphics[width=5.5in]{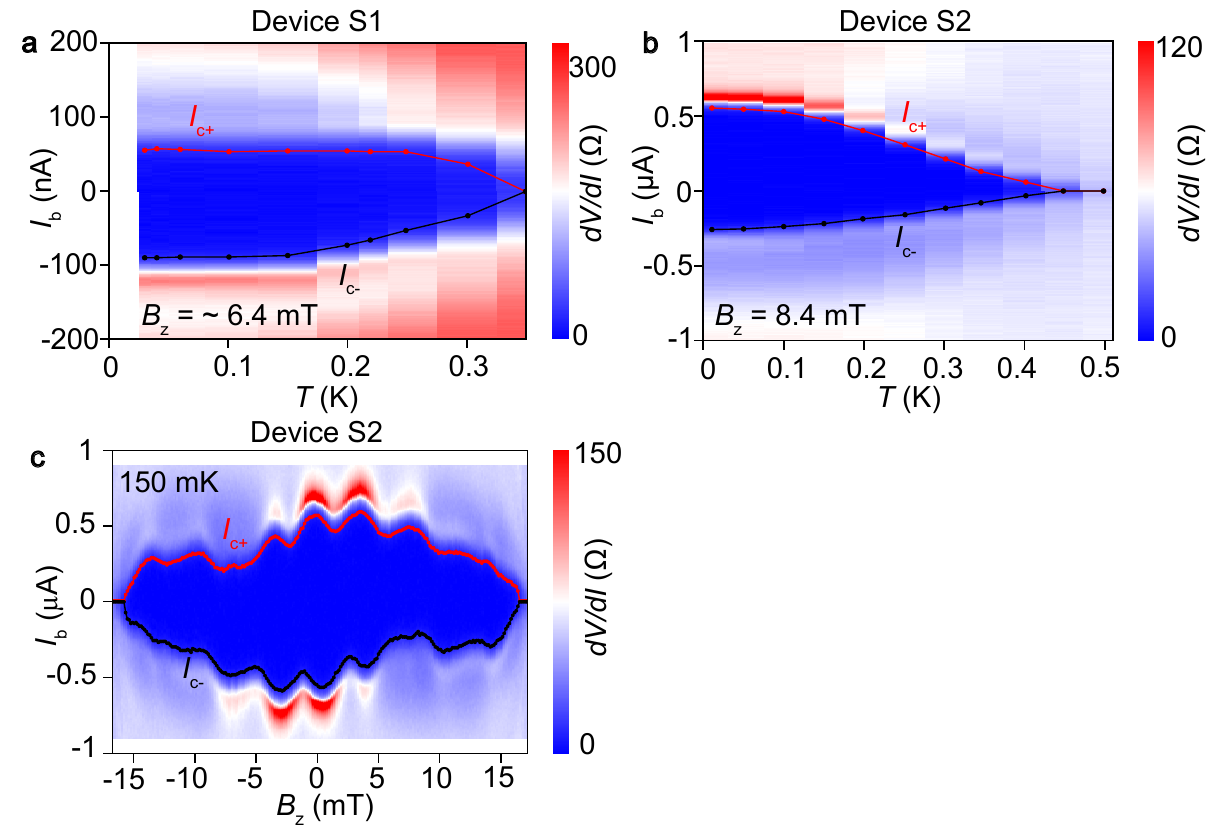}
\end{center}
\caption{\label{Fig2} \textbf{Temperature-dependent JDE for device S1 and S2.}
$\mathbf{a,b}$, Temperature-dependent JDE for device S1 and S2. $I_{c+}$ and $I_{c-}$ are marked by red and black scatters, respectively. $\mathbf{c}$, Interference pattern of device S2 at 150 mK.
  }
\end{figure*}

\begin{figure*}[!thb]
\begin{center}
\includegraphics[width=6in]{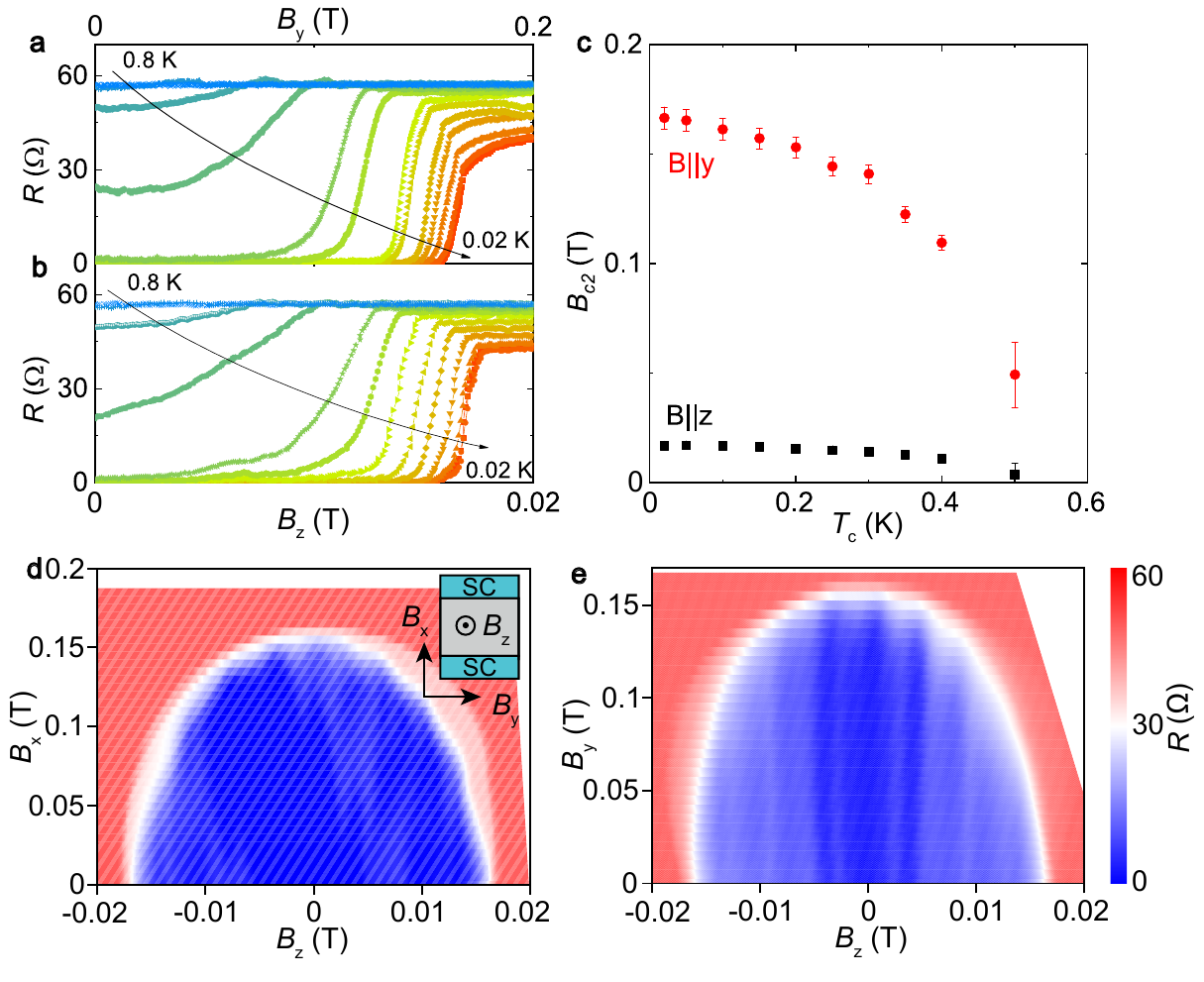}
\end{center}
\caption{\label{Fig2} \textbf{Magnetic field-dependent resistance for Ta$_{2}$Pd$_{3}$Te$_{5}$ device S2.}
$\mathbf{a, b}$, Superconducting transitions at different magnetic field and
$I_{ac}$ = 30 nA in device S2.
$\mathbf{c}$, Temperature-dependent upper critical field.
The error bars stem from the definition of the critical field,
which is determined by 50\% $\pm$ 10\% drop of the normal resistance.
$\mathbf{d}$, In-plane ($B_{x}$) and out-of-plane ($B_{z}$) magnetic field
dependence of resistance at $I_{ac}$ = 300 nA.
The inset shows the schematic diagram of the JJ, and the color bar shares the same with that in $\mathbf{e}$.
$\mathbf{e}$, $B_{z}$-dependent resistance at $I_{ac}$ = 400 nA and various $B_{y}$.
  }
\end{figure*}

\begin{figure*}[!thb]
\begin{center}
\includegraphics[width=6.5in]{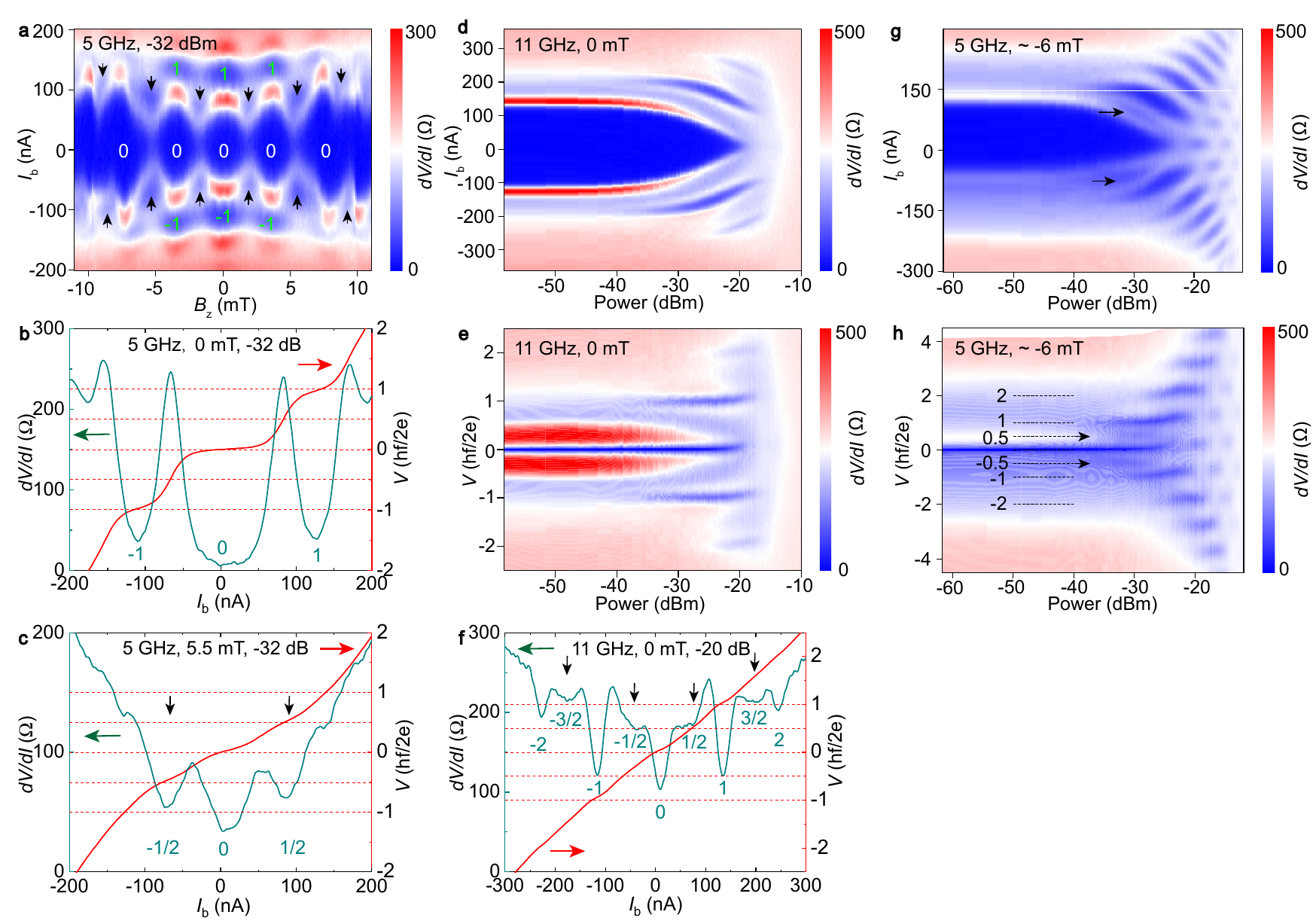}
\end{center}
\caption{\label{Fig5} \textbf{Shapiro steps as a function of $B_{z}$ in device S1.}
$\mathbf{a}$, Dependence of $dV/dI$ on $B_{z}$ and $I_{b}$
at microwave power = -32 dBm and frequency = 5 GHz.
The $\pm1/2$th Shapiro steps marked by black arrows exist at half flux quantum.
$\mathbf{b}$, Integer Shapiro steps observed in $I_{b}$-dependent $dV/dI$ and $V$ at 0 mT.
$\mathbf{c}$, Fractional Shapiro steps at 5.5 mT.
$\mathbf{d}$, Dependence of $dV/dI$ on microwave power and $I_{b}$ at 0 mT and 11 GHz.
$\mathbf{e}$, Shapiro steps of $\mathbf{d}$.
$\mathbf{f}$, Indistinctly visible half steps along the line cut at -20 dB in $\mathbf{e}$.
$\mathbf{g}$, Dependence of $dV/dI$ on microwave power at $\sim$ -6 mT and 5 GHz.
$\mathbf{h}$, Shapiro steps of $\mathbf{g}$. Half steps are observed.
  }
\end{figure*}

\begin{figure*}[!thb]
\begin{center}
\includegraphics[width=5.2in]{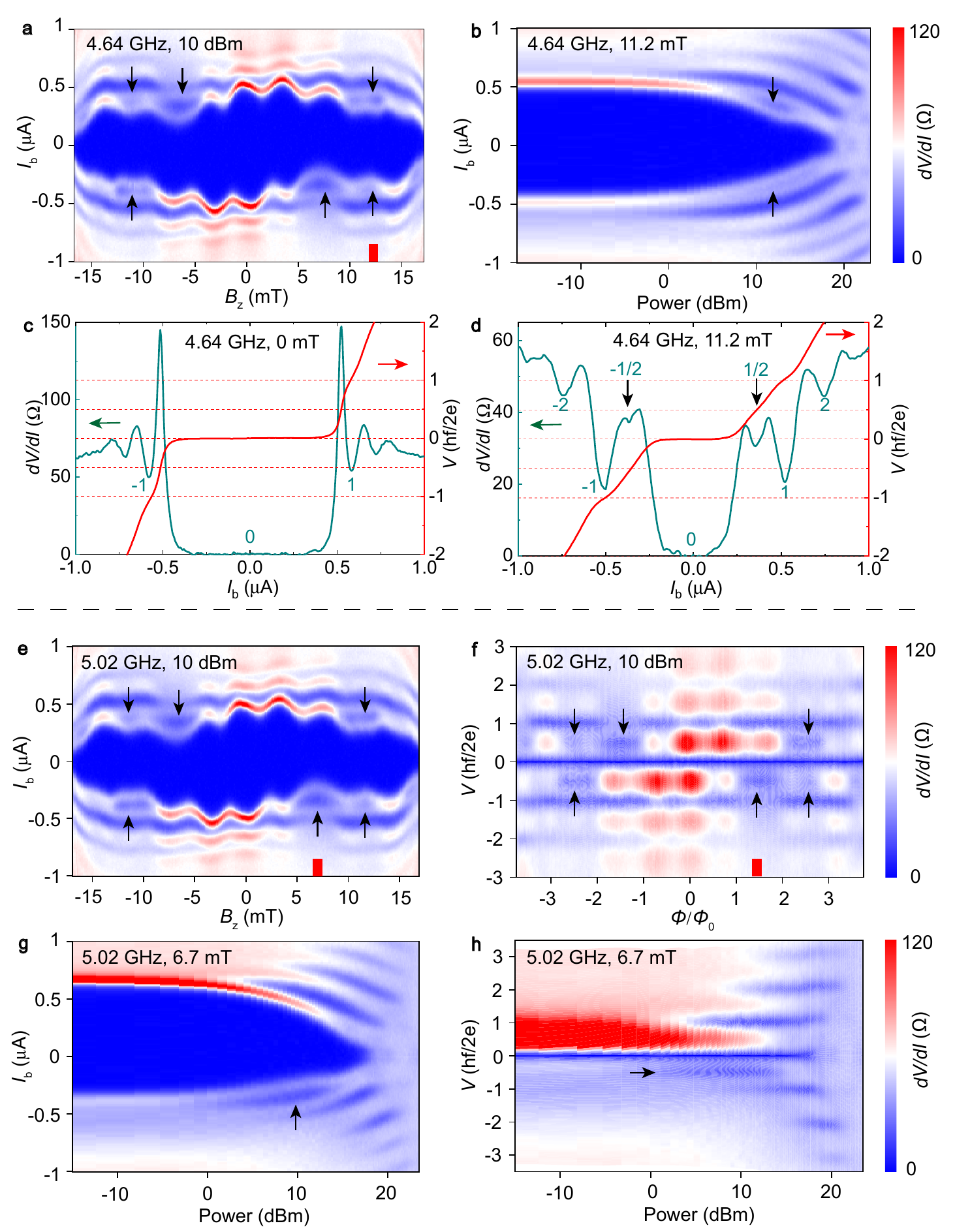}
\end{center}
\caption{\label{Fig10} \textbf{Fractional Shapiro steps in device S2.}
$\mathbf{a}$, $dV/dI$ as a function of $B_{z}$ and $I_{b}$
at microwave power = 10 dBm, 4.64 GHz and 10 mK.
The $\pm1/2$th Shapiro steps are marked by black arrows.
$\mathbf{b}$, Dependence of $dV/dI$ on microwave power and $I_{b}$ at 11.2 mT (5/2 $\phi_{0}$).
$\mathbf{c}$, $dV/dI$ - $I_{b}$ (cyan line) and $I_{b}$ - $V$ (red line) at $B_{z}$ = 0 mT.
$\mathbf{d}$, $dV/dI$ - $I_{b}$ (cyan line) and $I_{b}$ - $V$ (red line) at $B_{z}$ = 11.2 mT.
$\mathbf{e}$, $dV/dI$ as a function of $B_{z}$ and $I_{b}$
at microwave power = 10 dBm, 5.02 GHz and 10 mK.
$\mathbf{f}$, Shapiro steps versus flux quantum.
$\mathbf{g}$, Dependence of $dV/dI$ on microwave power and $I_{b}$ at 6.7 mT (3/2 $\phi_{0}$).
$\mathbf{h}$, Shapiro steps as a function of power.
$\mathbf{a}$, $\mathbf{e}$, $\mathbf{g}$ and $\mathbf{b}$, $\mathbf{f}$, $\mathbf{h}$ all share the same color bars.
  }
\end{figure*}

\vspace{3ex}

\noindent\textbf{Supplementary Note 4. Temperature-dependent JDE}

The temperature dependence of JDE under specific magnetic field is shown in Supplementary Figure 4 for both Ta$_{2}$Pd$_{3}$Te$_{5}$ devices.
The switching current $I_{c+}$ and $I_{c-}$, extracted by $\sim$ 15\% of normal resistance,
are plotted by red and black lines in Supplementary Figure 4a, b, respectively.
The obvious difference between $I_{c+}$ and $I_{c-}$ signifies the presence of the JDE.
The interference pattern at 150 mK in Supplementary Figure 4c includes all the oscillating features,
indicating the nearly all the interference features at 10 mK in Figure 2f of the main text.

\vspace{3ex}

\noindent\textbf{Supplementary Note 5. Excluding effects arising from finite-momentum superconductor}

Finite-momentum Cooper pairing is one of mechanisms responsible for inducing the superconducting diode effect (SDE)/JDE under an in-plane magnetic field $B_{y}$ \cite{SDEFMS_YuanNFQ_PNAS2022,NiTe2JDE_PalB_NP2022}.
Typically, the upper critical field ($B_{c2}$) under $B_{y}$ is greatly enhanced
in finite-momentum superconductors at ultra-low temperatures,
and it can exceed the Pauli limit ($B_{p}$ = 1.86 $T_{c}$) in thin films \cite{SDEFMS_YuanNFQ_PNAS2022}.
For device S2, $B_{c2}$ is $\sim$ 0.16 T under $B_{y}$ in Supplementary Figure 5c,
which is much smaller than the Pauli limit ($B_{p}$ = 0.95 T).
In addition, no enhancement of $B_{c2}$ appears at ultra-low temperature.
Additionally, the characteristic butterfly pattern attributed to finite-momentum Cooper pairing \cite{NiTe2JDE_PalB_NP2022} is not observed in Supplementary Figure 5d, e.
Therefore, it can be concluded that the $B_{z}$-induced JDE in Ta$_{2}$Pd$_{3}$Te$_{5}$ supercurrent interferometer does not originate from the finite-momentum Cooper pairing.

\vspace{3ex}

\noindent\textbf{Supplementary Note 6. Fractional Shapiro steps under microwave}

The Shapiro steps are usually measured to study the symmetry of Cooper pairing
and the higher harmonic contribution in the CPR \cite{4piPeriodJJ_NC2016,FST_Chauvin_PRL2006}.
Here we have performed measurements of fractional Shapiro steps to confirm the existence of the higher harmonic.
For device S1, $B_{z}$-dependent Shapiro steps are measured in Supplementary Figure 6a,
where the $\pm1/2$th Shapiro steps at $(n+1/2)\phi_{0}$ ($n$ is the integer) are marked by black arrows \cite{InAsAlSQUIDjde_arXiv2023,2DgasSQUIDdiode_arXiv2023}.
The $I_{b} - V$ curve at 0 mT in Supplementary Figure 6b demonstrates well-defined integer Shapiro steps.
At increased microwave frequencies (11 GHz in Supplementary Figure 6d, e, f),
fractional Shapiro steps are vaguely observed at 0 mT in our systems.
No higher frequency is further applied, considering the limits of our microwave generator.
Notably, conspicuous half-integer steps are also observed at $\sim$ -6 mT in Supplementary Figure 6e, f.

For device S2, $dV/dI$ as a function of $B_{z}$ at 5.02 GHz is also measured to check the presence of half-integer steps, as shown in Supplementary Figure 7e, f. The $\pm1/2$th Shapiro steps marked by black arrows exist at $(n+1/2)\phi_{0}$ (Supplementary Figure 7a,b, e-h), and are also detected between 0 and 15 dBm in Supplementary Figure 7b, g, h.

The robustness of $\pm1/2$th Shapiro steps at $(n+1/2)\phi_{0}$ is evident in device S1 (Supplementary Figure 6a) compared to device S2 in Supplementary Figure 7a, because the former exhibits one set of SQUID pattern.
For device S2, the total SQUID pattern exhibits the superposition of multiple SQUID patterns,
resulting in the different destructive interference of supercurrents compared with device S1.
Therefore, half-integer steps in both devices illustrate the existence of the second harmonic in CPR.

The step sizes of the $n$th Shapiro steps are subsequently studied.
In Supplementary Figure 8a, edges of the 0th and $\pm$1st Shapiro steps as a function of $B_{z}$ are
extracted from Supplementary Figure 7a and marked by $I_{c+}^{1st}$ (top edge of the 1st Shapiro step),
$I_{c+}^{0th}$ (top edge of the 0th Shapiro step or low edge of the 1st Shapiro step),
$I_{c-}^{0th}$ (low edge of the 0th Shapiro step or top edge of the 1st Shapiro step),
$I_{c-}^{1st}$ (low edge of the -1st Shapiro step).
The step sizes of the 0th and $\pm$1st Shapiro step are calculated by $I_{c+}^{0th} - \left|I_{c-}^{0th}\right|$,
$\left|I_{c+}^{1st} - I_{c+}^{0th}\right|$ and $\left|I_{c-}^{1st} - I_{c-}^{0th}\right|$, respectively.
As a result, in Supplementary Figure 8b, the antisymmetric behavior is observed
in the $B_{z}$-dependent difference between the 1st and -1st Shapiro step size
($\left|I_{c+}^{1st} - I_{c+}^{0th}\right|$ - $\left|I_{c-}^{1st} - I_{c-}^{0th}\right|$),
as well as that of the 0th Shapiro step ($I_{c+}^{0th} - \left|I_{c-}^{0th}\right|$),
consistent with the theoretical prediction \cite{FDEinSI_Constanti_PRL2022}.
To provide a clear view,
a line cut from Supplementary Figure 7a at $B_{z}$ = 2.9 mT is shown in Supplementary Figure 8c.
In addition, the power-dependent step sizes of the 0th and $\pm$1st Shapiro steps extracted from Figure 4e
of the main text are also displayed in Supplementary Figure 8d.
A difference between the $\pm$1st Shapiro step sizes (purple curve) is observed,
as well as the 0th Shapiro step sizes (orange curve).
Furthermore, the $\Delta I_{c}$ of the 0th (orange curve) and 1st Shapiro step (wine curve)
decreases slowly with increasing microwave power.
This stability of the JDE under microwave irradiation
presents potential applications in quantum information.

\textbf{Discussion of flux-dependent half-integer Shapiro steps.}
The visibility of fractional Shapiro steps is associated with the presence of higher order harmonics in the CPR.
Simply speaking, the disappearance of the first harmonic contribution will make fractional Shapiro steps more visible.
In other words, a sufficiently large spacing ($\Delta I_{dc}^{nosteps}$) is required between the consecutive integer Shapiro steps to observe fractional Shapiro steps \cite{FractionalShapiro_PRB2020}.
The $\Delta I_{dc}^{nosteps}$ is related to the reduced frequency $\omega = \phi_{0}f/2\pi RI_{c}$,
where $f$ is the microwave frequency. A larger value of $\omega$ leads to a larger $\Delta I_{dc}^{nosteps}$,
which in turn makes fractional Shapiro steps more evident.
In our work, $\omega$ in device S1 is 0.04 and 0.09 at 0 mT for $f$ = 5 GHz and 11 GHz, respectively,
and $\omega$ in device S2 is 0.04 for $f$ = 4.64 GHz at zero field.
The $\omega$ values do not seems large enough to observe half steps at zero field.
According to B. Raes $et$ $al.$ \cite{FractionalShapiro_PRB2020}, magnetic field is an effective method to increase $\Delta I_{dc}^{nosteps}$ by decreasing $I_{c}$, making half steps more visible. In device S1,
a noticeable decrease in $I_{c}$ at relatively small magnetic fields, due to the destructive interference of SQUID,
results in the appearance of half steps (Figure 3a of the main text, for example at 5.5 mT, where $\phi/\phi_{0}$ = 1.5,
$I_{c} \sim$ 20 nA, and $\omega$ = 0.26), providing insight into the existence of half steps at half flux.
To observe half steps at zero field in device S1, a higher frequency is required.
In device S2, the disappearance of half steps at $\phi/\phi_{0}$ = 0.5 may also be explained by the small $\omega$ = 0.04 (0.06) in $I_{c}^{+}$ ($I_{c}^{-}$) direction.
The disappearance (appearance) of half steps at $\phi/\phi_{0}$ = 1.5 in $I_{c}^{+}$ ($I_{c}^{-}$)
direction is attributed to the small $\omega$ = 0.05 (and a slightly larger $\omega \sim$ 0.1)
in the measurements in Supplementary Figure 7a.

\begin{figure*}[!thb]
\begin{center}
\includegraphics[width=5.5in]{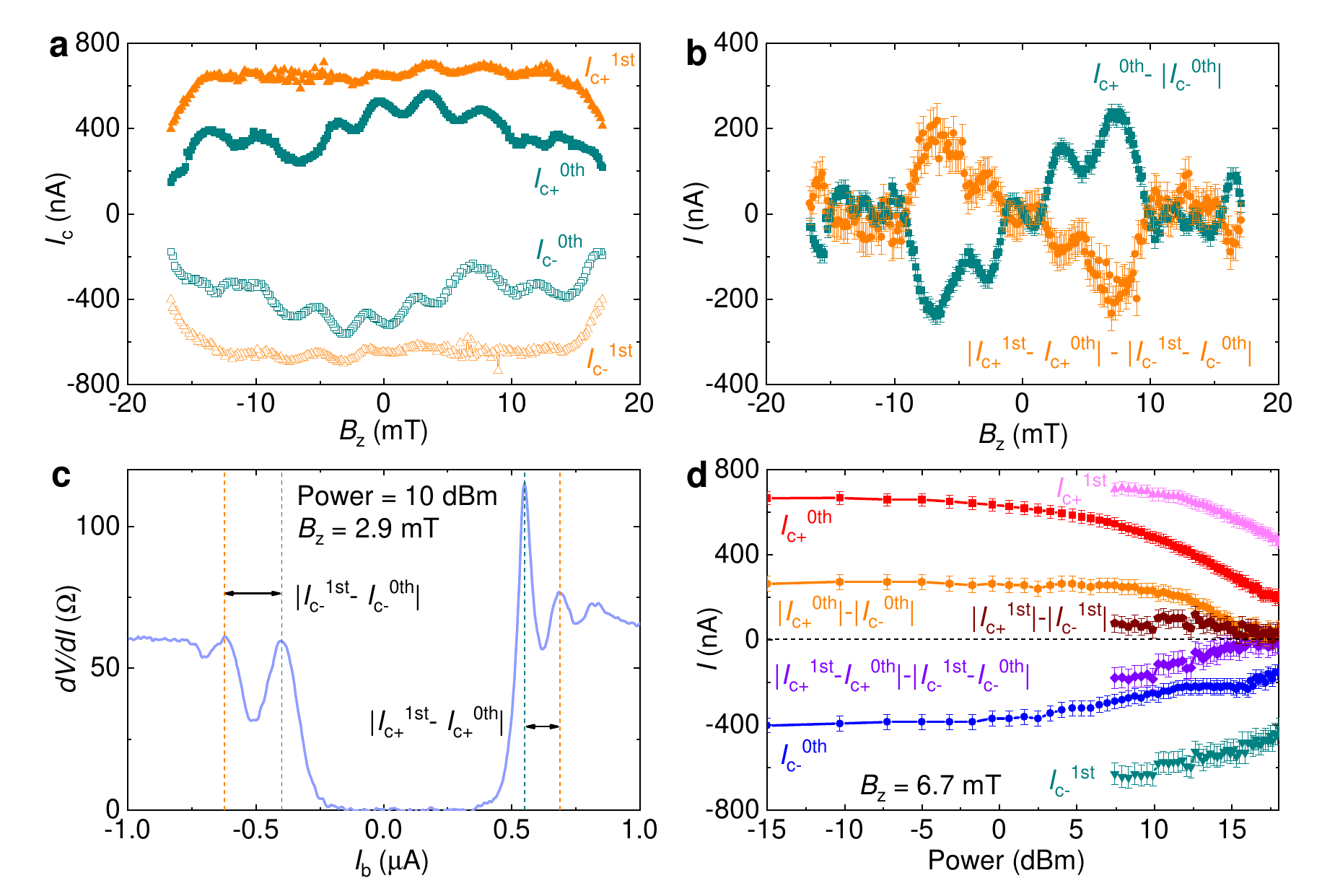}
\end{center}
\caption{\label{Fig5} \textbf{Shapiro steps sizes as a function of $B_{z}$ and microwave power in device S2.}
$\mathbf{a}$, Edges of the 0th and $\pm$1st Shapiro steps as a function of $B_{z}$.
When the JJ is subjected to microwave power,
the edges or switching currents can be determined by peaks in the $dV/dI$ - $I_b$ curve like $\mathbf{c}$.
$\mathbf{b}$, The difference of step sizes for the 0th and 1st Shapiro steps.
The error bars primarily originate from the definition of the switching current
due to the broadening of peaks in the $dV/dI$ - $I_b$ curve.
$\mathbf{c}$, $dV/dI - I_{b}$ at $B_{z}$ = 2.9 mT.
The step sizes between the 1st and -1st Shapiro steps are distinctly different.
$\mathbf{d}$, Microwave power-dependent Shapiro steps.
The error bars are obtained by the same method as in $\mathbf{b}$.
  }
\end{figure*}

\clearpage

\begin{figure*}[!thb]
\begin{center}
\includegraphics[width=5.5in]{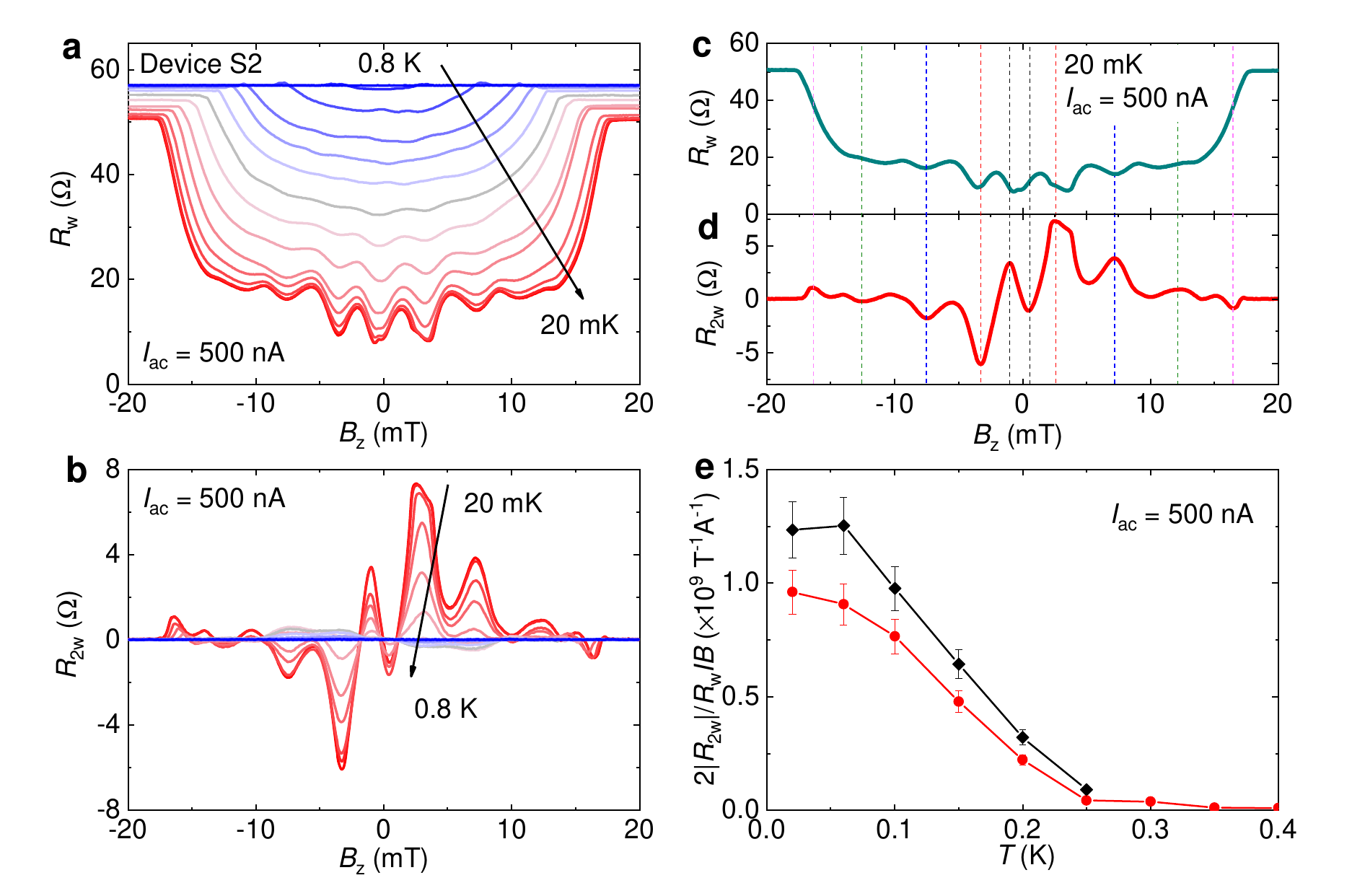}
\end{center}
\caption{\label{Fig11} \textbf{Temperature-dependent antisymmetric $R_{2w}$ of Ta$_{2}$Pd$_{3}$Te$_{5}$ device S2.}
$\mathbf{a}$, $R_{w}$ versus $B_{z}$ at different temperatures.
$\mathbf{b}$, Antisymmetric $R_{2w}$ with respect to $B_{z}$ at various temperatures.
$\mathbf{c,d}$, Typical $R_{w}$ and $R_{2w}$ curves at 500 nA.
Typical pairs of antisymmetric peaks ($R_{2w}$) are marked by dashed lines with the same colors.
$\mathbf{e}$, Temperature-dependent 2$\left| R_{2w}\right|/R_{w}IB$.
The average 2$\left| R_{2w}\right|/R_{w}IB$ or $\left| R_{2w}\right|$ is estimated using pairs of antisymmetric peaks in $\mathbf{d}$.
The error bars primarily originate from discrepancies between average $\left| R_{2w}\right|$ and the observed values.
  }
\end{figure*}

\noindent\textbf{Supplementary Note 7. Antisymmetric second harmonic transport of both devices}

The antisymmetric second harmonic transport ($R_{2w}$) typically occurs in the noncentrosymmetric or chiral systems \cite{MoS2NCT_WakatsukiR_SA2017,NbTaVSDE_AndoF_nature2020} and is usually associated with the magnetochiral anisotropy (MCA).
In noncentrosymmetric superconductors, such as superconductors with Rashba SOC or Ising SOC,
nonlinear responses under current $\mathbf{I}$ (electric field $\mathbf{E}$) and magnetic field ($\mathbf{B}$)
can be written as $R = R_{0} (1 + \gamma \mathbf{P} \times \mathbf{I} \cdot \mathbf{B})$
or $R = R_{0} (1 + \gamma IB)$ \cite{NonreciprocalTran_Tokura_NC2018},
where $\mathbf{P}$ is the inversion symmetry breaking direction and $\gamma$ is the MCA coefficient.
Rashba SOC (Ising SOC) $\mathbf{g} \propto \mathbf{P} \times \mathbf{I}$ \cite{SDEFMS_YuanNFQ_PNAS2022}.
Consequently, noncentrosymmetric superconductors generally exhibit antisymmetric peaks in $B$-dependent $R_{2w}$ curves.

In the supercurrent interferometer, the asymmetric nature of the device can also give rise to
antisymmetric $R_{2w}$ with respect to $B$.
Supplementary Figure 9b shows the antisymmetric oscillations in $R_{2w}$ as a function of magnetic fields.
In Supplementary Figure 9e, if we calculate 2$\left| R_{2w}\right|/R_{w}IB$ values using pairs of antisymmetric peaks or valleys
(black and red dashed lines in Supplementary Figure 9c, d), and the largest values can reach $1.2 \times 10^{9}$
and $9.6 \times 10^{8}$ T$^{-1}$A$^{-1}$, respectively.
However, it should be noted that the values of 2$\left| R_{2w}\right|/R_{w}IB$ estimated at large $I_{ac}$ and below the critical field can not correctly describe typical MCA or $\gamma$.
This is because the oscillating peaks of $\left| R_{2w}\right|$ as a function of $I_{ac}$ deviate from the origin of coordinates
(Figure 4c of the main text) and do not accord with common $R_{2w}=\gamma R_{w}BI$ formula.
The similar effect can also be observed in device S1, as shown in Supplementary Figure 10c.
Therefore, the MCA in the interferometer seems more complex and needs further investigation.

Therefore, a small $I_{ac}$ should be used to characterize the MCA,
with $\gamma$ calculated from a pair of typical antisymmetric peaks in the $R_{2w}$ - $B$ curves
near the critical fields ($\sim$ 16 mT).
For device S2 (Figure 4 of the main text), if the case of a 50 nA current is used to roughly estimate $\gamma$,
yielding a value of 7.0 $\times 10^{8}$ T$^{-1}$A$^{-1}$.
For device S1 (Supplementary Figure 10), if we roughly calculate $\gamma$ at $I_{ac}$ = 20 nA,
the value is approximately 3.12 $\times 10^{8}$ T$^{-1}$A$^{-1}$.
To compare with the MCA in several typical systems,
$\gamma' =\gamma A$, where $A$ is the cross-sectional area of the device,
is also calculated to characterize the MCA.
The $\gamma'$ of device S2 (device S1) can reach $4.92 \times 10^{-5}$ ($6.48 \times 10^{-6}$) m$^{2}$T$^{-1}$A$^{-1}$
with $A$ = $7.03 \times 10^{-14}$ ($2.08 \times 10^{-14}$) m$^{2}$.
These values are among the highest reported so far. \cite{BiSbTe3_LeggHF_NN2022,TdMoTe2SDE_arXiv2023}

\begin{figure*}[!thb]
\begin{center}
\includegraphics[width=6.5in]{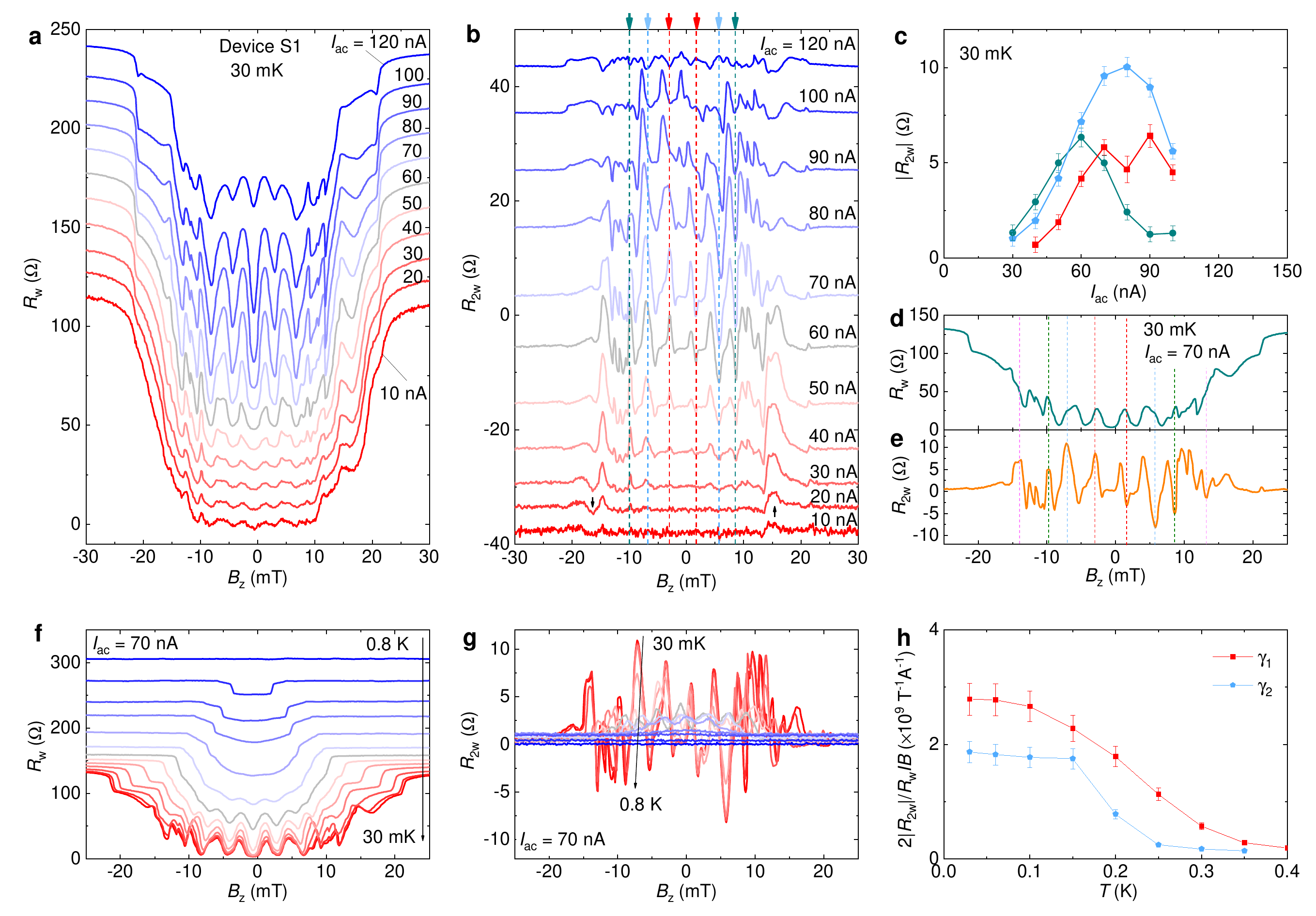}
\end{center}
\caption{\label{Fig13} \textbf{Current and temperature-dependent second harmonic transport of device S1.}
$\mathbf{a,b}$, $B_{z}$ dependence of $R_{w}$ and $R_{2w}$ at different $I_{ac}$.
Pairs of antisymmetric peaks ($R_{2w}$) are marked by dashed lines with the same colors.
The antisymmetric peaks at critical fields are marked by lack arrows at 20 nA.
$\mathbf{c}$, Current-dependent $\left| R_{2w}\right|$ for different oscillating peaks.
The error bars are obtained by the same method as in Supplementary Figure 9e.
$\mathbf{d,e}$, $R_{w}$ and $R_{2w}$ curves at 70 nA and 30 mK.
$\mathbf{f,g}$, Temperature-dependent $R_{w}$ and $R_{2w}$.
$\mathbf{h}$, Calculated 2$\left| R_{2w}\right|/R_{w}IB$.
The error bars are obtained by the same method as in Supplementary Figure 9e
  }
\end{figure*}

\clearpage

\noindent\textbf{Supplementary Note 8. Discussion on self-inductance of interferometer}

In this work, JDE is related to a non-sinusoidal CPR under a magnetic field,
but it could also be influenced by the screening effect, which has the potential to induce external magnetic flux in the superconducting loop \cite{tinkham2004introductiontoSC}.
To eliminate the effect of this self-generated flux, the self-inductance of the superconducting film is estimated,
including both geometric inductance ($L_{g}$) and kinetic inductance ($L_{k}$).
The geometric inductance $L_{g}$ is $\sim$ 2.2 pH (1.6 pH) for the equivalent rectangle
SQUID of device S2 (device S1) \cite{grover2004inductance}.
For the Al superconducting film deposited by our equipment, its normal resistivity at low temperature
is 0.375 $\mu\Omega\cdot cm$ with $T_{c} = 1.36$ K, as shown in Supplementary Figure 13.
The kinetic inductance $L_{k}$ is nearly 0.8 pH (0.6 pH) for device S2 (device S1)
according to the Eq. (6) in Anthony $et$ $al.$ \cite{KineticInductance_Nanotech2010}.
The critical current $I_{c}$ for device S2 (device S1) is $\sim$ 660 nA (126 nA) at zero magnetic field,
with $L\cdot I_{c}/\Phi_{0} \sim 10^{-3}$ ($\sim 1.3 \times 10^{-4}$) for device S2 (device S1)
\cite{SPM_WangJY_SA2022,SkewCPR_npjQM2020}.
Consequently, the self-inductance effect can be disregarded in our JJs.

\begin{figure*}[!thb]
\begin{center}
\includegraphics[width=4in]{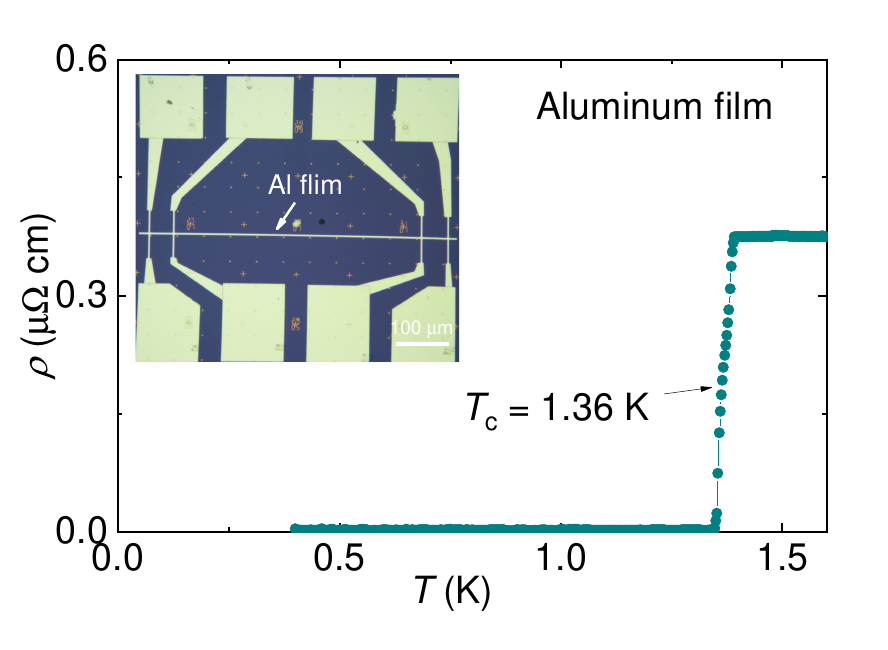}
\end{center}
\caption{\label{Fig4} \textbf{Temperature dependence of resistance in the Al film.}
The Al film deposited by our system hosts $T_{c}$ = 1.36 K.
The inset is the photomicrograph of the Al film, and the white scale bar corresponds to 100 $\mu$m.
  }
\end{figure*}

\noindent\textbf{Supplementary Note 9. Analysis of supercurrent density profile}

This section shows the Dynes-Fulton method \cite{DynesFulton_PRB1971,EdgeSQUID_np2014} for transferring the SQUID pattern $I_{c} (B)$
(red scatter points in Supplementary Figure 12a) to the supercurrent density profile $J_{c} (x)$ (Supplementary Figure 12d), using device S1 as an example.

Under a magnetic field, the total critical current flowing through the JJ is determined by a phase-sensitive summation of supercurrent across its width ($x$ diraction). Thus, the complex critical supercurrent function $\Im_{c} (\beta)$ can be described by:
\begin{equation}\label{LK}
\Im_{c} (\beta) = \int_{-\infty}^{\infty}dx J_{c} (x) e^{i\beta x},
\end{equation}
where the normalized magnetic field unit $\beta = 2\pi(L+L_{Al})B/\phi_{0}$, $L$ is the separation of the two electrodes, and $L_{Al}$ is the width of the Al electrode. Experimentally, the observed $I_{c} (B)$ represents the
magnitude of this summation: $I_{c} (\beta)$ = $\left| \Im_{c} (\beta) \right|$. Therefore, the $\Im_{c} (\beta)$ can be described as:
\begin{equation}\label{LK}
\Im_{c} (\beta) = I_{E} (\beta) + iI_{O} (\beta),
\end{equation}
where even [$I_{E} (\beta)$] and odd part [$I_{O} (\beta)$] are extracted from the $I_{c} (\beta)$.
The observed critical current $I_{c} (\beta)$ = $\sqrt(I_{E}^{2} (\beta) + I_{O}^{2} (\beta))$.
The even part of the critical current $I_{E} (\beta)$ represents a symmetric distribution and
is given by $I_{E} (B) = \int_{-\infty}^{\infty}J_{E}(x) cos(\beta x)dx$.
This can be roughly obtained by multiplying $I_{c} (\beta)$ with a flipping function (see Supplementary Figure 12b).
Odd part of the critical current is given by $I_{O} (B) = \int_{-\infty}^{\infty}J_{O}(x) sin(\beta x)dx$,
which can be approximated by interpolating between the minima of $I_{c}^{max}$,
and flipping sign between lobes (Supplementary Figure 12c).
Therefore, the supercurrent density profile (Supplementary Figure 12d) is:
\begin{equation}\label{LK}
J_{c} (x) = \left| \frac{1}{2\pi} \int_{-b/2}^{b/2}\Im_{c}(\beta) e^{-i\beta x} d\beta \right|,
\end{equation}
where $b$ is the sampling range of $\beta$.

\begin{figure*}[!thb]
\begin{center}
\includegraphics[width=6in]{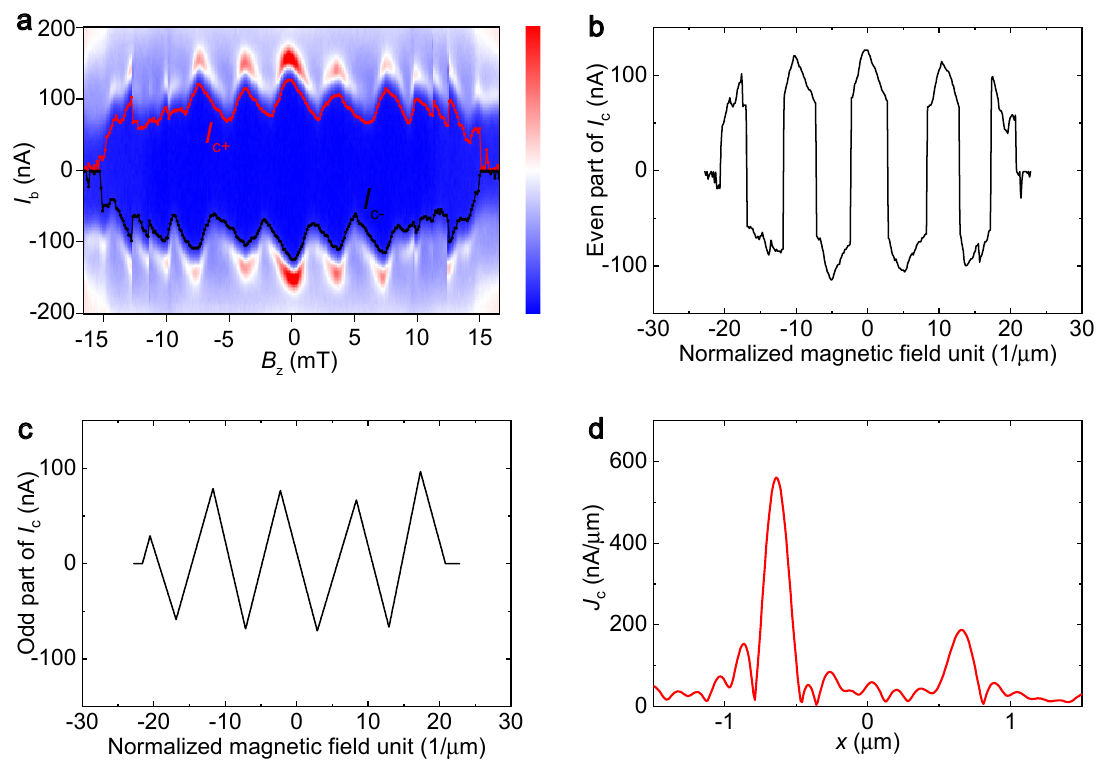}
\end{center}
\caption{\label{Fig4} \textbf{Analysis of current density distribution for device S1.}
$\mathbf{a}$, The whole SQUID pattern of device S1 at 15 mK. $I_{c+}$ and $I_{c-}$ are marked by red and black scatters, respectively.
The interference pattern beyond the range of $\pm$10 mT appears somewhat irregular,
potentially due to factors such as vortex pinning close to critical magnetic field.
$\mathbf{b}$, The critical current $I_{E}$ that corresponds to the even part of the current
density profile $J_{E}$.
$\mathbf{c}$, The critical current $I_{O}$ that corresponds to the even part of the current
density profile $J_{O}$.
$\mathbf{d}$, Current density distribution of device S1.}
\end{figure*}

\clearpage
